\documentclass[floatfix, preprint, showpacs, showkeys, preprintnumbers, nofootinbib, superscriptaddress]{revtex4-1}
\usepackage[utf8]{inputenc}
\usepackage[sort&compress]{natbib}
\usepackage{ulem}
\usepackage{bm}
\usepackage{times}
\usepackage{amssymb,amsbsy,amsmath,amsfonts}
\usepackage{graphicx}
\usepackage{float}
\usepackage{color}
\usepackage{morefloats}
\usepackage{rotating}
\usepackage{srcltx}
\usepackage{slashed}
\usepackage{subfigure}
\usepackage{multirow}
\usepackage{verbatim}
\usepackage{hyperref}
\usepackage{tabularx}

\newcommand{\RNum}[1]{\uppercase\expandafter{\romannumeral #1\relax}}

\begin{document}

\title{Nuclear charge radii in Bayesian neural networks revisited }
\author{Xiao-Xu Dong}
\affiliation{School of Physics,  Beihang University, Beijing 102206, China}

\author{Rong An}
\affiliation{Key Laboratory of Beam Technology of Ministry of Education, Institute of Radiation Technology, Beijing Academy of Science and Technology, Beijing 100875, China}
\affiliation{Key Laboratory of Beam Technology of Ministry of Education, College of Nuclear Science and Technology, Beijing Normal University, Beijing 100875, China}
\affiliation{CAS Key Laboratory of High Precision Nuclear Spectroscopy, Institute of Modern Physics, Chinese Academy of Sciences, Lanzhou 730000, China}

\author{Jun-Xu Lu}
\email[E-mail: ]{ljxwohool@buaa.edu.cn}
\affiliation{School of Space and Environment,  Beihang University, Beijing 102206, China}
\affiliation{School of Physics,  Beihang University, Beijing 102206, China}

\author{Li-Sheng Geng}
\email[E-mail: ]{lisheng.geng@buaa.edu.cn}
\affiliation{School of
Physics,  Beihang University, Beijing 102206, China}
\affiliation{Beijing Key Laboratory of Advanced Nuclear Materials and Physics, Beihang University, Beijing 102206, China }
\affiliation{School of Physics and Microelectronics, Zhengzhou University, Zhengzhou, Henan 450001, China }

\begin{abstract}

In this work, a refined Bayesian neural network (BNN) based approach with six inputs including the proton number, mass number,   and engineered features associated with the pairing effect, shell effect, isospin effect, and  ``abnormal" shape staggering effect of $^{181,183,185}$Hg, is proposed to accurately describe  nuclear charge radii.
The new approach is able to well describe the charge radii of atomic nuclei with $A\ge40$ and $Z\ge20$. The standard root-mean-square (rms) deviation is  $0.014$ fm for both the training and validation data.  In particular,  the predicted charge radii of proton-rich and neutron-rich calcium isotopes are found in good agreement with data. We further demonstrate the reliability of the BNN approach by investigating the variations of the rms deviation with extrapolation distances,   mass numbers, and  isospin asymmetries.

\end{abstract}


\maketitle

\section{Introduction}
Nuclear charge radius is one of the most fundamental properties of an atomic nucleus, which characterizes its charge distribution. It is a key observable that can directly  reflect various fine structure phenomena, such as neutron halo~\cite{Nortershauser:2008vp,Geithner:2008zz}, neutron skin~\cite{Brown:2017xxo,Yang:2017vih}, odd-even staggering~\cite{GarciaRuiz:2016ohj,Miller2019,An:2021yoj}, shape staggering~\cite{Marsh:2018wxs,Barzakh:2021gfl}, and the emergence  of new magic numbers~\cite{Angeli:2015wia}. Remarkable experimental progress has been achieved in measuring charge radii over the past few years. Particularly, laser spectroscopy experiments have measured more than one hundred charge radii of unstable nuclei~\cite{Angeli:2013epw,Li:2021fmk,Barzakh:2021gfl,Pineda:2021shy,Malbrunot-Ettenauer:2021fnr,Reponen:2021rwy}. Some exotic and interesting phenomena, such as  the endpoint of the shape staggering of mercury isotopes~\cite{Marsh:2018wxs} and the abrupt increase  in the charge radii of neutron-rich calcium isotopes ~\cite{GarciaRuiz:2016ohj} as well as the odd-even staggering in proton-rich calcium isotopes~\cite{Miller2019}, pose great challenges to our  understanding of nuclear charge radii.  Therefore a systematical study of the new experimental data is of great importance.

Various theoretical models have been applied to study nuclear charge radii, from simple liquid drop models~\cite{Weizsacker:1935bkz,Brown:1984zz}, phenomenological formulae~\cite{Nerlo-Pomorska:1994dhg,Zhang:2001nt,Wang:2013zia,Sheng:2015poa,Wang:2013zia,Li:2021fmk}, local-relation based models~\cite{Peng:2014jia,Bao:2016suw,Sun:2016nec,Bao:2020ffv,Ma:2021jzu}, sophisticated mean-field models~\cite{Geng:2005yu,Goriely:2016sdz,Pena-Arteaga:2016clz,Sarriguren:2019jfb,An:2020qgp,An:2021rlw,DRHBcMassTable:2022uhi} to  $ab$ $initio$ no core shell models~\cite{Forssen:2009vu}. Among these models, the Weizs$\ddot{\mathrm{a}}$cker Skyrme~(WS$^{*}$) model is able to provide the best description of the experimental data, yielding a root-mean-square deviation (RMSD) around 0.018 fm~\cite{Li:2021fmk}. However, the description and explanation of some fine structures, such as the odd-even staggering of calcium isotopes, still remain difficult for most theoretical models. In 2020, a modified relativistic mean field plus BCS (RMF(BCS)*) ansatz~\cite{An:2020qgp}, which considers the semi-microscopic correction originating from the Cooper pair condensation, was proposed to describe the charge radii of  calcium isotopes, and  the agreement with data turns out to be quite good, compatible with or even slightly better than the sophisticated Fayans energy density functional approach~\cite{Reinhard:2017ugx,Miller2019}. 

In recent years, machine learning methods have been widely and successfully applied to study various physical systems~\cite{Carleo:2019ptp,Bourilkov:2019yoi,Bedolla-Montiel:2020rio,Bedaque:2021bja,Boehnlein:2021eym}. In particular, a variety of machine learning methods are used to study nuclear charge radii, ranging from the naive Bayesian probability classifier~\cite{Ma:2020rdk}, kernel ridge regression (KRR) model~\cite{Ma:2022yrj}, artificial neural networks~(ANNs)~\cite{Akkoyun:2012yf,Wu:2020bao,Yang:2022tjf}, to Bayesian neural networks~(BNNs)~\cite{Utama:2016tcl,Dong:2021aqg}. In Ref.~\cite{Dong:2021aqg}, a hybrid model combining the simplicity of a three-parameter formula (NP) with the expression power of a Bayesian neural network, named as the D4 model, was proposed to describe nuclear charge radii with $Z\ge20$ and $A\ge40$. It can not only achieve a much improved description of experimental charge radii~\cite{Angeli:2013epw,Goodacre:2020sys,Li:2021fmk} but also can describe the peculiar odd-even staggering of calcium isotopes and make predictions for neutron-rich calcium and potassium isotopes with quantified uncertainties, in good agreement with the experimental data~\cite{Koszorus:2020mgn,Li:2021fmk}. Nonetheless, a closer examination of the predictions shows that the D4 model does not give satisfactory predictions for isotopes in the  neutron-deficient region of several isotopic chains, i.e.,  calcium and thallium.  In addition, the rms deviation for the validation set is about 30\% larger than that for the training set, which hints at a possible over-fitting problem.
In this work,  we refine the NP-BNN4 (D4) model of Ref.~\cite{Dong:2021aqg} by adding new features containing more physical information, trying to provide a solution to the problems mentioned above and develop a new NP-BNN model with better extrapolation ability.

This article is organized as follows. In Sec. II, we construct the refined Bayesian neural network by adding new features and explain how we divide experimental charge radii into training and validation sets. Results and discussions are presented in Sec. III. A short summary and outlook is provided in Sec. IV.

\section{Theoretical Formalism}

\begin{figure}[htbp]
    \centering
    \includegraphics[angle=270,width=0.8\textwidth]{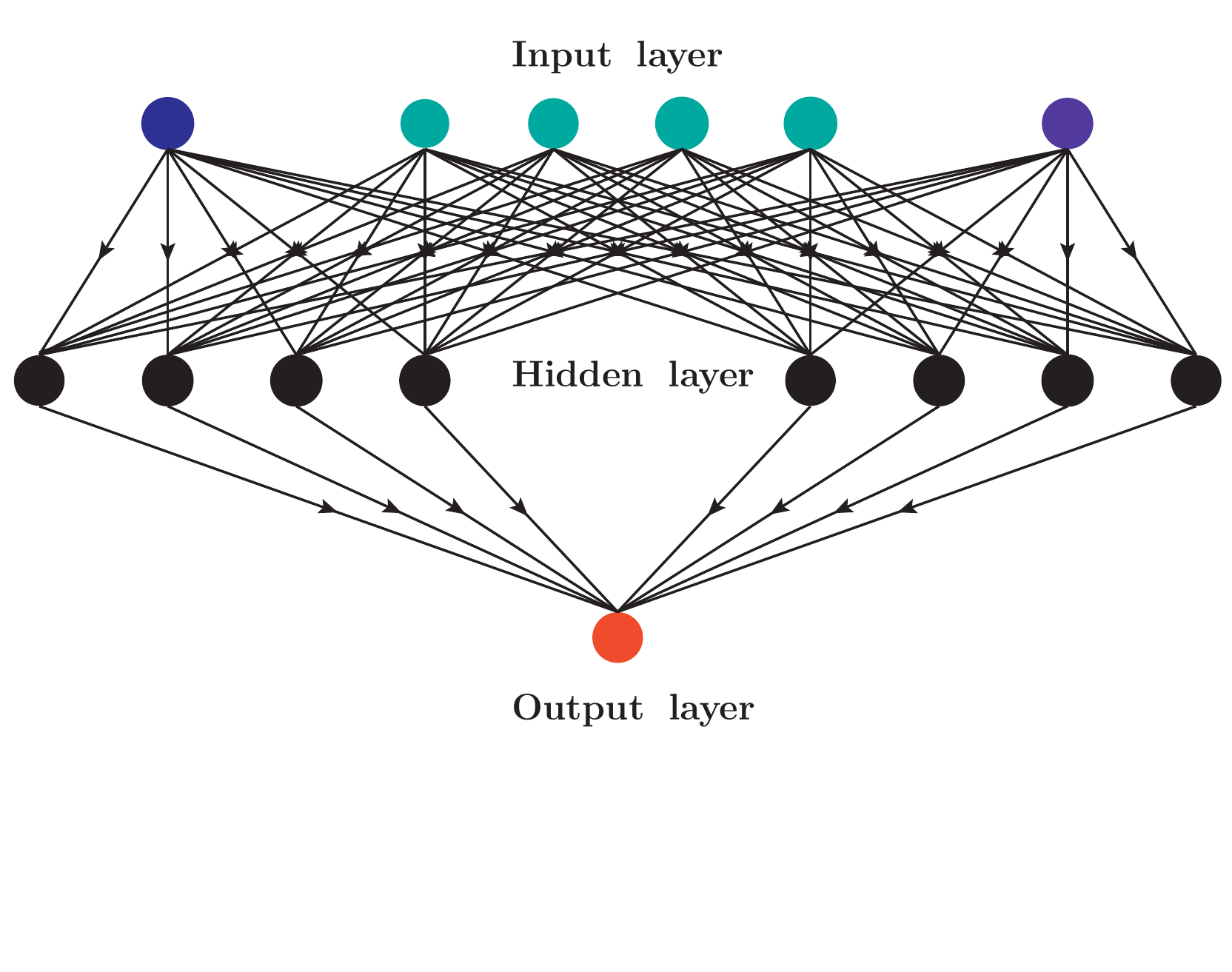}
    \caption{ 
    Structure of the  artificial neural network used in this work. The number of neurons in the input layer is 6. The number of hidden layers is 1 and the number of neurons in the hidden layer is $34$. The number of neurons in the output layer is $1$.
    } 
    \label{Structure}
\end{figure}

The Bayesian neural network (BNN)~\cite{Neal} is a powerful machine learning method, because of its ability to combine the strengths of an artificial neural network (ANN) and the Bayesian statistical theory, with the former being generally regarded as a ``universal approximator''~\cite{HORNIK1989359}. As shown in Fig.~\ref{Structure}, the artificial neural network we used is a fully connected feed-forward artificial neural network with one hidden layer. Mathematically, it has the following form:
\begin{align}
&f(x,\omega)=a+\sum\limits_{j=1}\limits^H b_j \tanh(c_j+\sum\limits_{i=1}\limits^I d_{ji}x_i),\label{f1}
\end{align}
where the parameters of the neural network are $\omega=({a, b_j, c_j, d_{ji}})$, $I$ is the number of input layer neurons, $H$ is the number of  hidden layer neurons, and $x$ is the set of inputs $x_i$. 

The Bayesian inference is based on Bayes's theorem~\cite{Neal,James},
\begin{align}
p(\omega|x,t)=\frac{p(\omega)p(x,t|\omega)}{p(x,t)},
\end{align}
where $p(\omega)$ is the prior distribution of the parameters of the neural network, $p(x,t|\omega)$ is the likelihood based on the actual data,  $p(\omega|x,t)$ is the posterior distribution which is used to make predictions, $p(x,t)$ is the marginal likelihood which can always be neglected because  it does not contain the information of parameters, and $t$ is the set of target data $t_i$. All the model parameters are assumed to be independent in this work and follow Gaussian distributions centered around zero as their prior distributions. To define the likelihood $p(x,t|\omega)$, we first need to introduce an objective function for the neural network in terms of a least-squares fit to the training data:
\begin{align}
\chi^2(\omega)=\sum\limits_{i=1}\limits^N\left(\frac{t_i-f(x_i,\omega)}{\Delta t_i}\right)^2,
\end{align}
where $N$ is the number of training data. In this work, we use the experimental uncertainties of charge radii as $\Delta t$, the set of standard deviation $\Delta t_i$.
A Gaussian distribution is usually used for the likelihood in terms of the objective function:
\begin{align}
p(x,t|\omega)=\exp(-\frac{\chi^2(\omega)}{2}).
\end{align}

All that remain for the model training process are to specify the input data and target data of the Bayesian neural network. As we mentioned in the Introduction, the predictions of the D4 model for the neutron-deficient calcium isotopes do not agree well with the data~\cite{Dong:2021aqg}. The same can be said about the potassium isotopic chain. Clearly, isospin dependence is not adequately treated in the D4 model. In addition, the D4 model is not able to predict well the latest experimental data for neutron-deficient mercury and thallium isotopes. Experimentally, the charge radii of $^{181,182,183,184,185}$Hg show strong odd-even staggering, which has been attributed to the shape oscillation from prolate to oblate and back forth~\cite{Marsh:2018wxs}. In the D4 model, such shape staggering, which happens much less frequently, has not been explicitly considered and is not correctly captured by the network either.  To prevent such a rare event from distorting the BNN, we add a new feature by labeling  $^{181,183,185}$Hg such that they are treated differently from all the other nuclei. This should be viewed as the most economic way to remove ``abnormal'' data from the training set. 

Thus, the final inputs for the BNN are $x\equiv (Z,A,\delta,P,I^2,LI)$, where
\begin{align}
\delta&=\frac{(-1)^Z+(-1)^N}{2},\\
P&=\frac{\nu_p\nu_n}{\nu_p+\nu_n},\\
I^2&=(\frac{N-Z}{A})^2,\\
LI&=\left\{  
\begin{array}{c c}
1,&(Z,A)=(80,181),(80,183),(80,185)\\
0,&\mathrm{else}
\end{array}. \right.
\end{align}
 The pairing term $\delta$ is related to  nuclear pairing effects and the promiscuity factor $P$~\cite{Casten:1987zz,Casten_1996} is related to shell closure effects. The new features are $I^2$, which takes into account isospin dependence, and $LI$, which treats the ``abnormal'' behavior in $^{181,183,185}$Hg. It should be noted that we do not observe any such large shape oscillations in other nuclei except for  $^{188}$Bi~\cite{Barzakh:2021gfl}.
 The target data $t$ in the present work are $\delta R_{\mathrm{ch}}=R_{\mathrm{exp.}}-R_{\mathrm{th}}$, i.e., the residuals between experimental  and theoretical charge radii. Following Ref.~\cite{Dong:2021aqg}, we choose the NP formula~\cite{Nerlo-Pomorska:1994dhg} as our theoretical model to be refined by BNN:
\begin{align}
R_{\mathrm{NP}}(Z,A)=r_A A^{\frac{1}{3}}\left[ 1-b(\frac{N-Z}{A})+\frac{c}{A}\right],
\end{align}
where $r_A$=0.966 fm, $b=0.182$, and $c=1.652$~\cite{Bayram:2013jua}. 

The Bayesian prediction for the charge radius difference of a particular nucleus $n$ is:
\begin{align}
\langle f_n\rangle=\int f(x_n,\omega)p(\omega|x,t)d\omega=\frac{1}{K}\sum\limits_{k=1}\limits^K f(x_n,\omega_{k}),\label{fn}
\end{align}
where  $x_n=(Z_n,A_n,\delta_n,P_n,I^2_n,LI_n)$ are the input data (features), $f(x_n,\omega_{k})$ are the neural network predictions for $\delta R_{\mathrm{ch}}(Z_n,A_n)$ for a given set of parameters $\omega_{k}$, and $K$  is the total number of effective samples. In this work, we use the Markov chain Monte Carlo (MCMC)  method~\cite{Neal}  to obtain the Bayesian predictions. One of the most important features of BNNs (in comparison with conventional neural networks) is that they can provide a proper estimate of output uncertainties. To characterize such uncertainties in a quantitative way, we define a confidence interval of 68.3\% as :
\begin{align}
\Delta=\sqrt{\langle f_n^2 \rangle-\langle f_n \rangle^2},\label{delta}
\end{align}
where  $\langle f_n^2 \rangle $ is obtained by following the same procedure as in obtaining $\langle f_n\rangle $.

\section{Results and discussions}

Similar to \cite{Dong:2021aqg}, we only study those nuclei with $A\ge40$ and $Z\ge20$. For the training set, we use the 820 experimental data given in Ref.~\cite{Angeli:2013epw}. The more recent experimental data ~\cite{Goodacre:2020sys,Li:2021fmk}, containing 113 data for nuclei with $A\ge40$ and $Z\ge20$, are used as the validation set to test the  predictive power  of our model. 

For the sake of easy reference, we use  ``D6'' to denote  the model combining the NP formula and the BNN with 6 input neurons.  We also show the results obtained in the BNN with five input neurons $ (Z,A,\delta,P,I^2)$, denoted by ``D5'' for comparison. The difference between ``D5'' and ``D6'' shows how a local feature affects the global performance of a BNN via the large non-linearity of a neural network.  
To quantify the extent of the BNN refinement of the NP formula, we compute the root-mean-square deviation (RMSD)  between the NP-BNN model outputs $R^{(\mathrm{NB})}$ and experimental data $R^{(\mathrm{exp})}$:
\begin{equation}
\sigma^{(v)}(R^{(\mathrm{exp})},R^{(\mathrm{NB})})=\sqrt{\frac{1}{N_v}\sum\limits_{i=1}\limits^{N_v}\left(R_{i}^{(\mathrm{exp})}-R_{i}^{(\mathrm{NB})}\right)^2}\label{rmsd}
\end{equation}
where $N_v$ is the total number of charge radii contained in the validation set. 

One advantage of the BNN method is that it can provide  quantitative uncertainty estimates. However, the above conventional RMSD does not reflect this important piece of information. As a result, we propose to use a modified root mean square deviation named as special root mean square deviation (SRMSD), such that the generic model uncertainties are taken into account, which reads
\begin{align}
&S\sigma^{(v)}(R^{(\mathrm{exp})},R^{(\mathrm{NB})})=\sqrt{\frac{1}{N_v}\sum\limits_{i=1}\limits^{N_v}(S\sigma_i)^2},\label{Srmsd}\\
&S\sigma_i=\left\{  
\begin{array}{c c}
0,&\mathrm{if} R_i^{(L)}\le R_i^{\mathrm{(exp)}}\le R_i^{(H)}\\
|R_i^{\mathrm{(exp)}}-R_i^{(H)}|,&\mathrm{if} R_i^{\mathrm{(exp)}}>R_i^{(H)}\\
|R_i^{\mathrm{(exp)}}-R_i^{(L)}|,&\mathrm{if} R_i^{\mathrm{(exp)}}<R_i^{(L)}\\
\end{array} \right.\label{Srmsdi}
\end{align}
where $R_i^{(L)}=R^{\mathrm{(NB)}}-\Delta$ and $R_i^{(H)}=R^{\mathrm{(NB)}}+\Delta$, $R_i^{\mathrm{(NB)}}$ is the output of the NP-BNN model (D4/D5/D6) for nucleus $i$, and $\Delta$ is the confidence interval of Eq.~\eqref{delta}.

As can be seen from Eqs.~\eqref{rmsd}, \eqref{Srmsd} and \eqref{Srmsdi}, the SRMSD  reflects the deviations between the experimental data and the 1$\sigma$ limits of  theoretical predictions while the RMSD denotes the deviations between the experimental data and the center values of theoretical predictions.

  The RMSDs of the NP formula, D4~\cite{Dong:2021aqg}, D5, and D6 models are displayed in Table~\ref{rms} for the training set~($\sigma^{(t)}$) and the validation set~($\sigma^{(v)}$), where the SRMSDs for the validation set , $S\sigma^{(v)}$, are also shown.

\begin{table}[htpb]
\caption{RMSDs and SRMSDs for the charge radii predicted by the NP formula, D4, D5 and D6 models.}
\begin{center}
\begin{tabular}{c c c c }
  \hline\hline
  ~~Model~~ & ~~$\sigma^{(t)}~(\mathrm{fm})$~~ &~~$\sigma^{(v)}~(\mathrm{fm})$~~& ~~$S\sigma^{(v)}~(\mathrm{fm})$~~\\
  \hline
NP&0.0394 &0.0300   &\\
D4 &0.0143  & 0.0187  &0.0142\\
D5& 0.0137 & 0.0170  &0.0124\\
D6&0.0140  & 0.0139  &0.0094\\
  \hline\hline

\end{tabular}
\label{rms}
\end{center}
\end{table}

First, we note that all the three NP-BNN models can describe the training set much better than the NP formula, at a level of 0.014 fm. On the other hand, for the validation set, the D6 model yields the least RMSD, which is only about 46\% of that  of the NP formula, 74\% of the D4 model, and 82\% of the D5 model. In addition, the RMSD of the D6 model for the validation set is almost the same as that for the training set while that of the D4(D5) model increases by  31\%(24\%). This means that the D6 model performs much better in terms of extrapolation and its predictions are more reliable. 

It is interesting to compare the D5 and D6 models, which differ by the input feature ``$LI$''. The improvement of D6 over D5 indicates that a few ``abnormal'' data could distort the calibration of a neural network and thus affect its performance. Because of the scarcity of ``abnormal`` data, they could not be correctly identified by the neural network. In such a scenario, it is important to treat them separately by hand, if feasible, such that the distortion of the neural network performance could be minimized or eliminated.   

In the last column of Table~\ref{rms}, we show the SRMSDs for the D4/D5/D6 models. It is clear that taking into account uncertainties, the difference between theoretical  and experimental results  decreases by about 24\% for the D4 model, 27\% for the D5 model, and 32\% for the D6 model.

\begin{figure}[htbp]
\centering
\subfigure{
\begin{minipage}[t]{0.49\linewidth}
\centering
\includegraphics[width=1.0\textwidth]{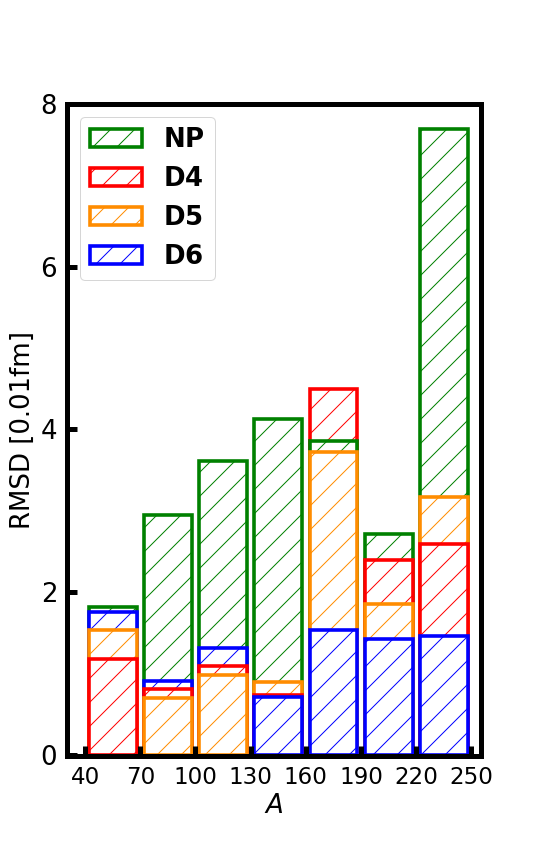}
\end{minipage}%
}
\subfigure{
\begin{minipage}[t]{0.49\linewidth}
\centering
\includegraphics[width=1.0\textwidth]{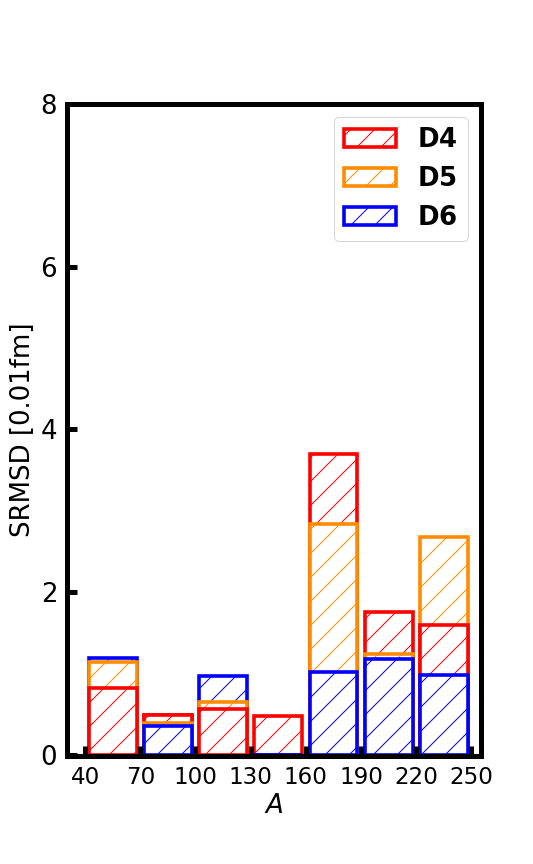}
\end{minipage}%
}%
\centering
\caption{Variations of the RMSDs and SRMSDs for nuclei with mass number $A$ in the validation set. }
\label{A}
\end{figure}

\begin{figure}[htbp]
\centering
\subfigure{
\begin{minipage}[t]{0.49\linewidth}
\centering
\includegraphics[width=1.0\textwidth]{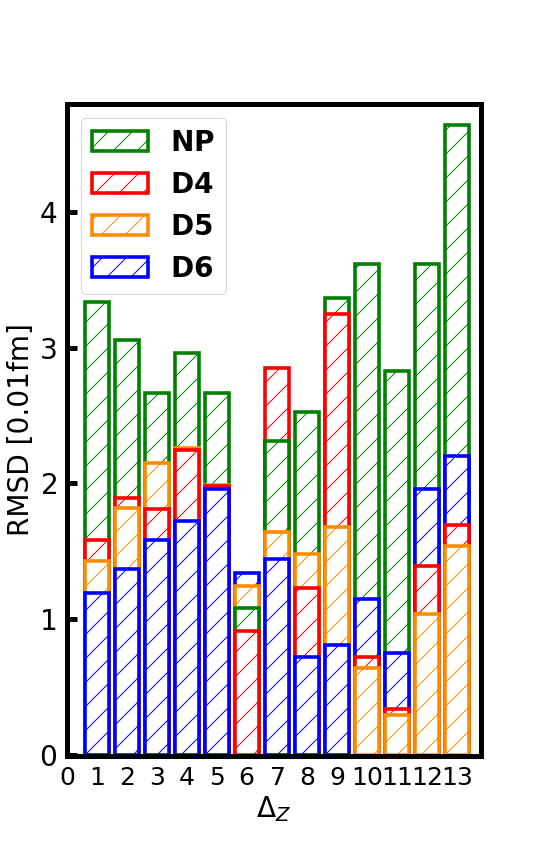}
\end{minipage}%
}%
\subfigure{
\begin{minipage}[t]{0.49\linewidth}
\centering
\includegraphics[width=1.0\textwidth]{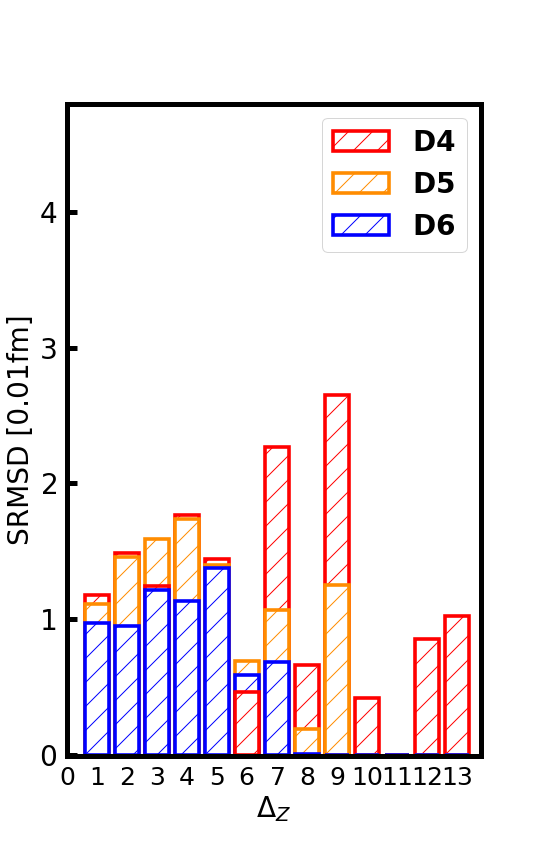}
\end{minipage}%
}%
\centering
\caption{Variations of RMSDs and SRMSDs for nuclei in the validation set with the extrapolation distance $\Delta_Z$. }
\label{EXD}
\end{figure}

\begin{figure}[htbp]
\centering
\subfigure{
\begin{minipage}[t]{0.49\linewidth}
\centering
\includegraphics[width=1.0\textwidth]{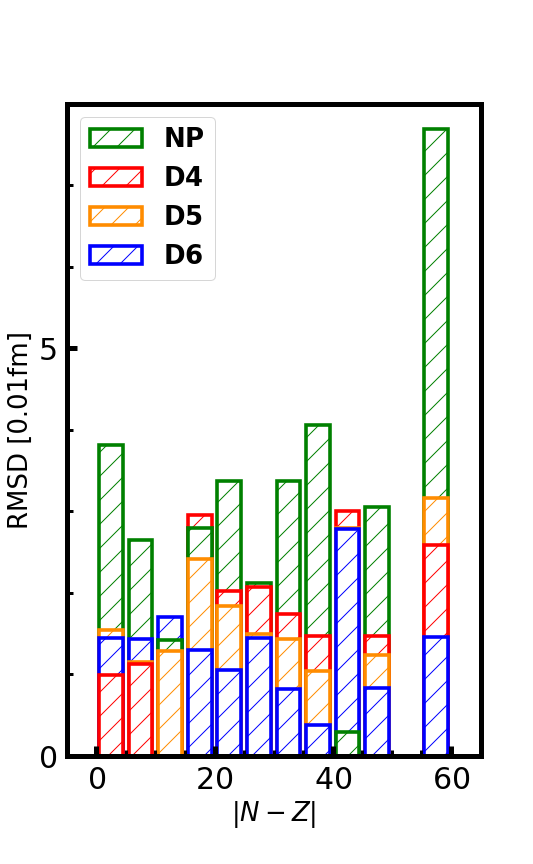}
\end{minipage}%
}%
\subfigure{
\begin{minipage}[t]{0.49\linewidth}
\centering
\includegraphics[width=1.0\textwidth]{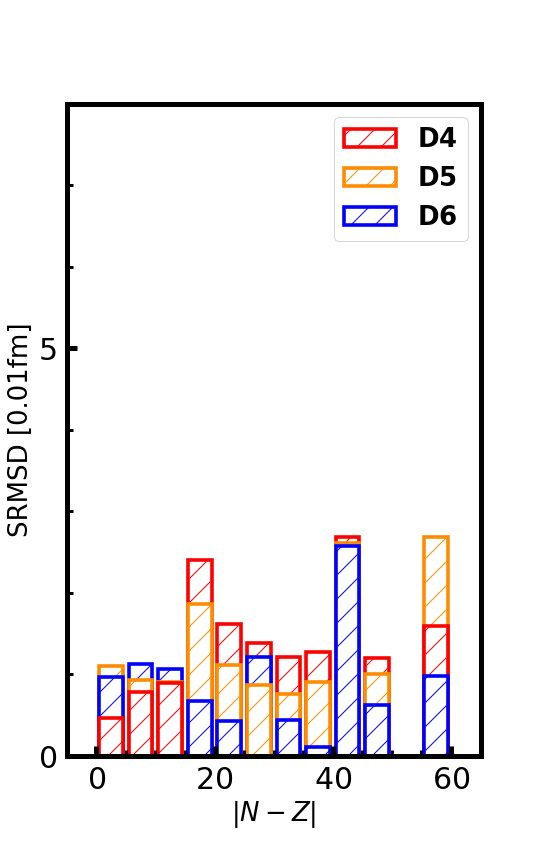}
\end{minipage}%
}%
\centering
\caption{Variation of the RMSDs and SRMSDs for the validation set with $|N-Z|$. Every bin in each interval represents the (S)RMSD of all the nuclei in this interval. }\label{NZ}
\end{figure}

A physically motivated and constrained NP-BNN model is supposed to perform better in extrapolations in terms of masses,  extrapolation distances, and isospin asymmetries.  Therefore, in the following,  we test the generalization ability of the NP-BNN models by studying the variations of RMSDs(SRMSDs)  with the mass number $A$, the extrapolation distance $\Delta_Z$, and the isospin asymmetry $|N-Z|$ respectively. For those nuclei in the same mass region, with the same $\Delta_Z$  and in the same $|N-Z|$ region, one can define a RMSD similar to Eq.~\eqref{rmsd} as well as a SRMSD similar to Eq.~\eqref{Srmsd}. The corresponding results are shown in Fig.~\ref{A}, Fig.~\ref{EXD}, and  Fig.~\ref{NZ}.

As can be seen from Fig.~\ref{A}, the RMSDs and SRMSDs of the D6 model exhibit less fluctuation compared with those of the D4  and D5 models. Especially in the heavy mass region, the performance of the D6 model is more stable, which implies that the D6 model is able to yield more reliable predictions for super-heavy nuclei, which is very important because of the lack of experimental data in this region.

To further check the predictive power of the NP-BNN models, we define the extrapolation distance $\Delta_Z$ for data in the validation set as follows:
\begin{align}
    &\Delta_Z(Z_i^{(v)},N_i^{(v)})=\mathop{\mathrm{MIN}}\limits_{Z_j^{(t)}=Z_i^{(v)}}\left(|N_j^{(t)}-N_i^{(v)}|\right)
\end{align}
where the subscripts $i$ and  $v$ denote a nucleus in the validation set, and $j$ and $t$ denote a nucleus in the training set. As a result, $\Delta_Z$ represents the shortest distance between the nuclei contained in the validation set and those in the training set with same proton number $Z$. 

The variations of the RMSDs and SRMSDs for  nuclei in the validation set with $\Delta_Z$ are shown in Fig.~\ref{EXD}. The RMSDs of the D6 model are smaller than those of the  D4 and D5 models for $\Delta_Z\leq 4$, which contains 64.6\% of the validation data.  Large odd-even staggering of the D4 predictions happens in the $5\leq \Delta_Z \leq 10$ region, which contains 31.9\% of the validation  data, while no large odd-even staggering  is found in the D6 predictions. Even with the  uncertainties taken into account, the large odd-even staggering behavior of the D4 model still persists. The origin of this odd-even staggering effect can be understood by studying the thallium isotopes. 
\begin{figure}[htbp]
\subfigure{
\begin{minipage}[t]{0.49\linewidth}
\centering
\includegraphics[width=1.0\textwidth]{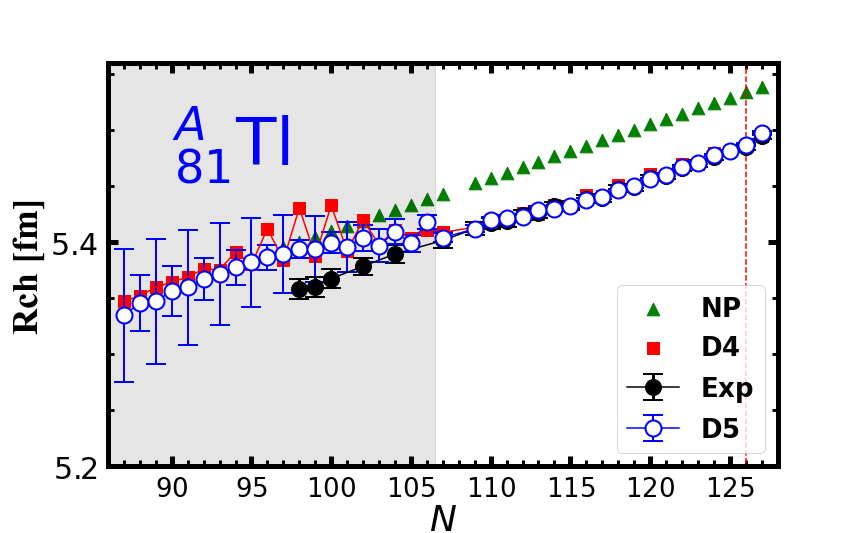}
\end{minipage}%
}%
\subfigure{
\begin{minipage}[t]{0.49\linewidth}
\centering
\includegraphics[width=1.0\textwidth]{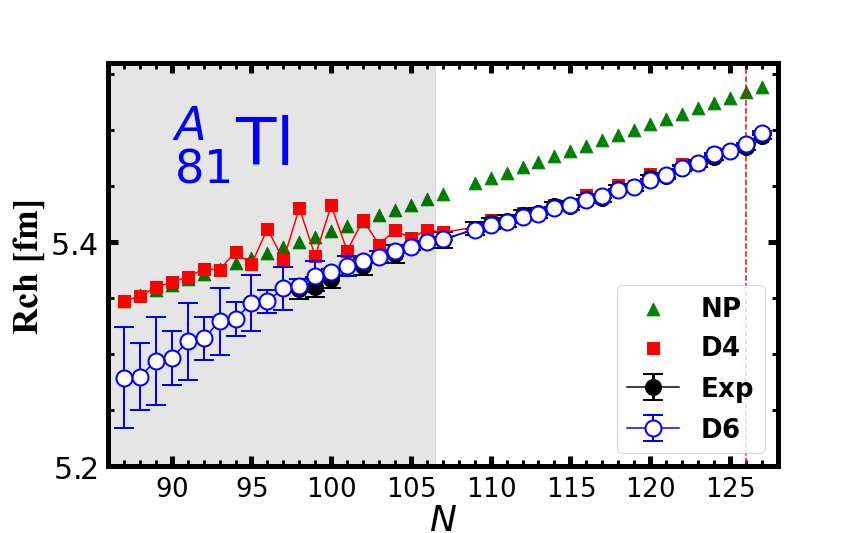}
\end{minipage}%
}%
\centering
\caption{Charge radii of the thallium isotopes predicted by the  NP formula~\cite{Nerlo-Pomorska:1994dhg,Bayram:2013jua}, D4 (center value)~\cite{Dong:2021aqg}, D5, and D6 models, in comparison with the experimental 
data~\cite{Angeli:2013epw,Li:2021fmk}. The data in the grey area are pure predictions, i.e., they are not contained in the training set. The red dashed line denotes the  neutron magic number $N=126$.}
\label{thallium}
\end{figure}
As can be seen in Fig.~\ref{thallium}, the predictions of the D4 model show large odd-even staggerings in conflict with  the experimental data. 
Though the new feature of isospin effect is not able to solve this problem, it decreases the amplitude of the odd-even staggerings as can be seen by comparing  the results of the D5 and D4 models. After the local interaction feature $LI$ is included, the predictions of the D6 model for the thallium isotopes are in excellent agreement with the experimental data.

Another interesting thing is that the RMSDs of the D6 model are not  monotonically increasing  with $\Delta_Z$ as those of the kernel ridge regression method~\cite{Ma:2022yrj}, which means that the extrapolations of the D6 model are more stable. After the BNN uncertainties are taken into account, the extrapolation capacity of the D6 model is further improved, especially for those nuclei far from those contained the training set. All the experimental data of the validation set in the $\Delta_Z\geq 9$ region (about 11.5\% of the validation set ) are within the confidence intervals  provided by the D6 model.

In Fig.~\ref{NZ}, we show the variation of the RMSDs and SRMSDs of the validation set as a function of the distance to the  $N=Z$ line in the nuclear chart, to further investigate the effect of the new feature $I^2$.  With $I^2$ considered in the BNN,  both the RMSDs and SRMSDs decrease in  the $15\le |N-Z|\le 50$ region. In addition,  the RMSDs  of the D6 model are smaller than those of the D5 model in the $|N-Z|\ge 15$ region. As a result, it implies that the local feature introduced to account for $^{181,183,185}$Hg can influence the study of the nuclear charge radii in the whole nuclear chart via the complicated neural network.

\begin{figure}[htbp]
\centering
\includegraphics[width=0.48\textwidth]{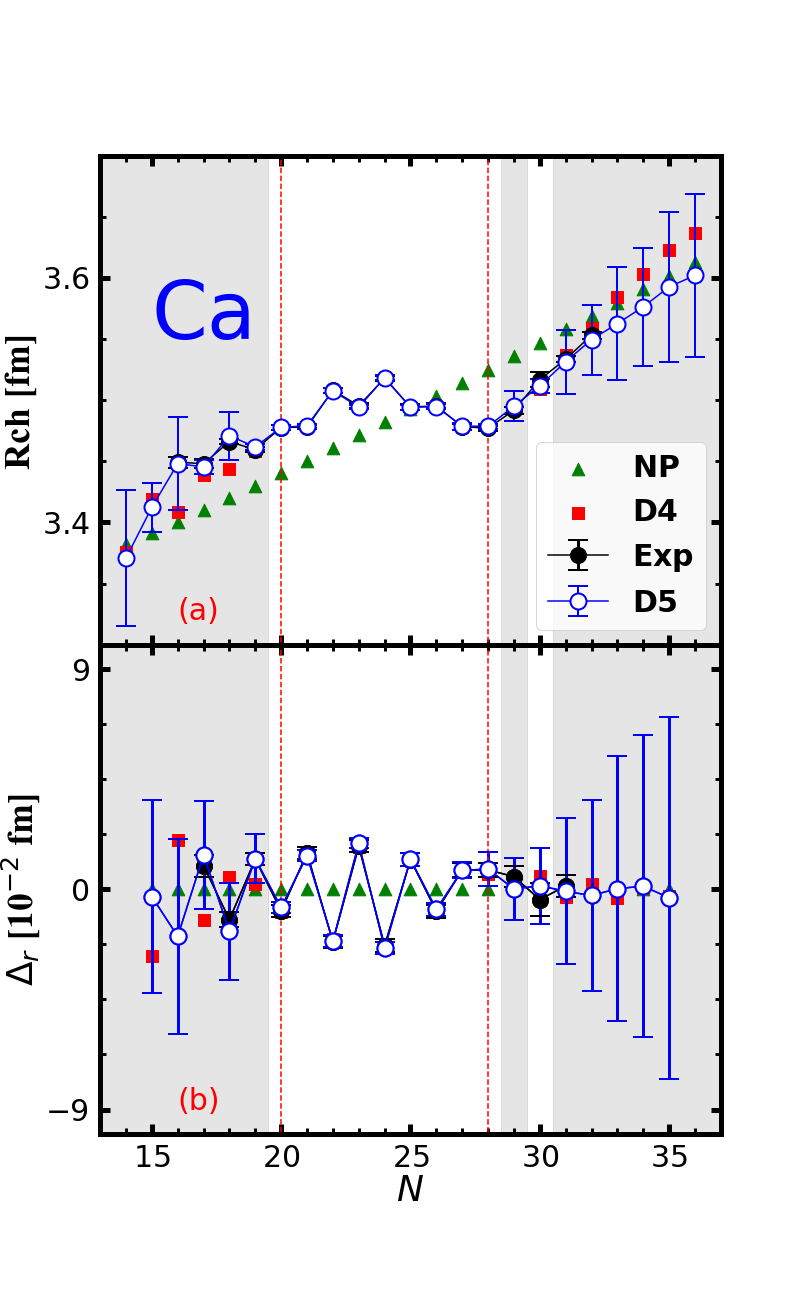}
\includegraphics[width=0.48\textwidth]{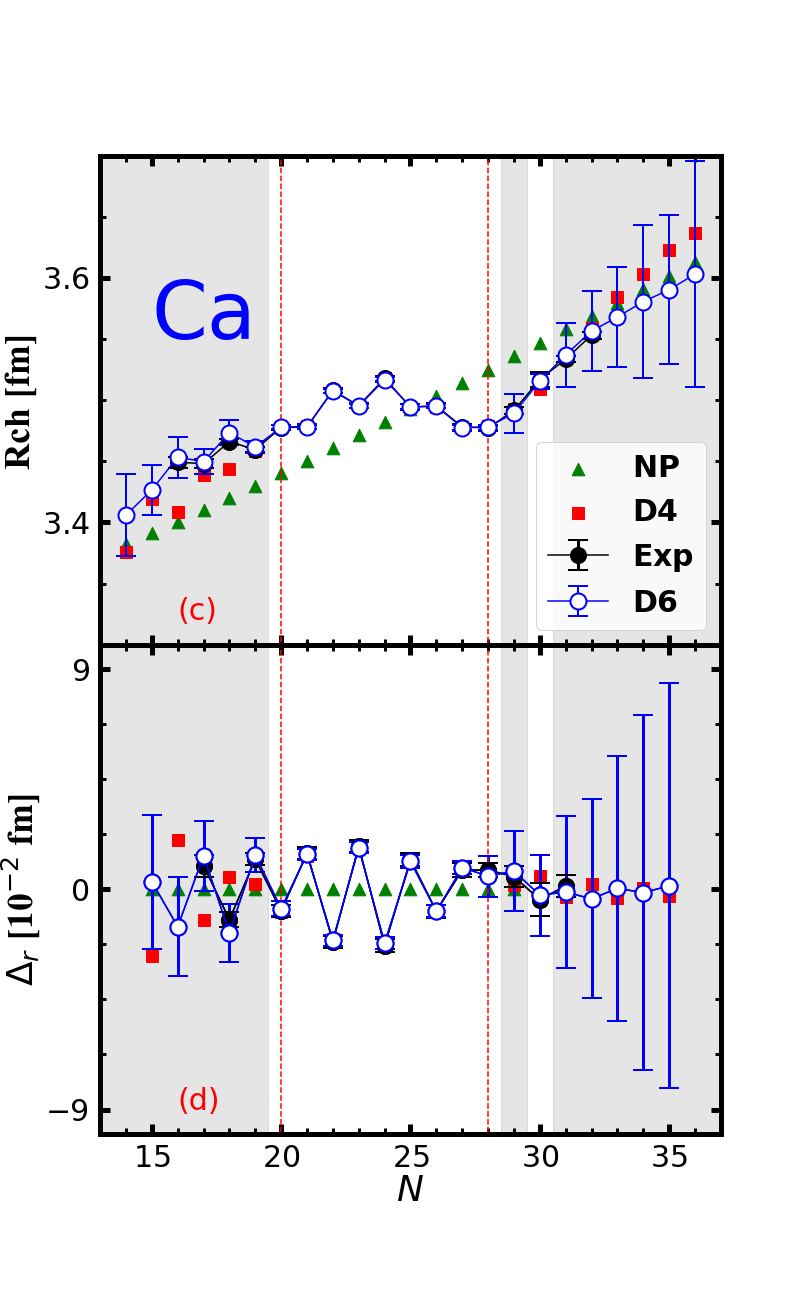}
\\
\centering
\caption{Charge radii (a, c) and $\Delta_r$ (b, d) of calcium isotopes predicted by the  NP formula~\cite{Nerlo-Pomorska:1994dhg,Bayram:2013jua}, D4  (center value), D5,   and D6 models, in comparison with the experimental data~\cite{Angeli:2013epw,Li:2021fmk}. The data in the grey area are pure predictions, i.e., they are not contained in the training set. The red dashed line represents the  neutron magic number 20 and 28. }\label{calcium}
\end{figure}

Clearly, from the above analysis, one can see that the predictive power of the D6 model in the $A\geq 40$ region is quite good. It is interesting to check whether in the $A<40$ region, such as the neutron-deficient calcium isotopes,  the D6 model  can still provide good predictions. In order to showcase the odd-even staggering effect of the calcium isotopes, we  study
the odd-even staggering $\Delta_r$ defined by
\begin{align}
    \Delta_r(N,Z)&=\frac{1}{2}\left[ R(N-1,Z)-2R(N,Z)+R(N+1,Z)\right],
\end{align}
where $R(N,Z)$ is the RMS charge radius for a nucleus with neutron number $N$ and proton number $Z$. 
The results are shown in the lower panels of Fig.~\ref{calcium}. As can be seen from Fig.~\ref{calcium}, with four input neurons ($Z,A,\delta,P)$, the predictions of the D4 model do not agree well with data for those nuclei with $N<19$. On the other hand, the D5 and D6 predictions are in better agreement with the experiment data in the $N<19$ region, as well as in the neutron-rich region. This can be viewed as a clear evidence that the isospin-asymmetry effect plays an important role in describing the charge radii and the odd-even staggering phenomena of the calcium isotopes, especially in the neutron-deficient region. On the other hand,  in the $29\le N \le 31$ region, the D6 model describes better the odd-even staggering than the D5 model, which further proves the importance of the explicit consideration of  local ``anomolies''  in global nuclear charge radii studies.

Finally, we briefly comment on the performance of various machine learning methods in studies of nuclear charge radii, particularly in describing the data in the validation set. For the sake of simplicity, we use  RMSDs for this purpose with the caveat that the nuclei contained in the validation sets can be very different.   By randomly dividing all the data~\cite{Angeli:2013epw} into a training set and a validation set, the  RMSDs achieved by the ANNs of Refs.~\cite{Akkoyun:2012yf,Wu:2020bao} are  0.023 fm. The naive Bayesian probability classifier achieves a RMSD of  0.02 fm~\cite{Ma:2020rdk} while the KRR method obtained a RMSD of  0.03 fm~\cite{Ma:2022yrj}. The  BNN methods of Ref.~\cite{Utama:2016tcl} and our previous study~\cite{Dong:2021aqg} achieved RMSDs of  0.026 fm and 0.019 fm, respectively. While in the present work, we have achieved a RMSD of  0.014 fm.

\section{Summary and outlook}

We showed that with  four physically motivated engineered features, i.e., isospin-asymmetry, pairing, shell and local shape-staggering effects, one can achieve an unprecedented description of nuclear charge radii. Compared to the three-parameter parameterization, the RMSD(SRMSD) of the validation set achieved by the NP-BNN6 model is lower by about 54\%(69\%).

Studying  the variation of RMSDs and SRMSDs  with nuclear mass number $A$, we found that the new features $I^2$ and $LI$ improved the predictive power of the NP-BNN model mainly in the heavy mass region. The large odd-even staggering effects of RMSDs and SRMSDs disappear after adding the new  features $I^2$ and $LI$ into the BNN. In addition, the predictive power of the D6 model becomes more apparent  when the theoretical uncertainties are taken into account. Even  for the nuclei located far from those contained in the training set, the D6 model can still make fair predictions.

The present work demonstrated the potential of Bayesian neural networks in explaining nuclear structure properties, such as  nuclear charge radii. In particular, we showed that physically motivated features are indispensable for the cases where data are limited and extrapolations are necessary. In addition, the power of Bayesian neural networks in providing quantitative uncertainties should also be fully exploited. The local feature considered showed that a local ``anomaly'', which indicates physics not explicitly considered, can have the potential to influence the descriptions and predictions elsewhere via the complicated neural network and should be taken care in similar studies.

\section{Acknowledgments}
This work was partly supported by the National Natural Science Foundation of China (NSFC) under Grants No. 11975041, No. 11735003, No. 12105006, and No. 11961141004. R. A. is supported in part by the Key Laboratory of High Precision Nuclear Spectroscopy, Institute of Modern Physics, Chinese Academy of Sciences. Junxu Lu acknowledges support from the National Natural Science Foundation of China under Grant No.12105006 and China Postdoctoral Science Foundation under Grant No. 2021M690008.

\bibliography{BNN.bib}

\begin{thebibliography}{55}%
\makeatletter
\providecommand \@ifxundefined [1]{%
 \@ifx{#1\undefined}
}%
\providecommand \@ifnum [1]{%
 \ifnum #1\expandafter \@firstoftwo
 \else \expandafter \@secondoftwo
 \fi
}%
\providecommand \@ifx [1]{%
 \ifx #1\expandafter \@firstoftwo
 \else \expandafter \@secondoftwo
 \fi
}%
\providecommand \natexlab [1]{#1}%
\providecommand \enquote  [1]{``#1''}%
\providecommand \bibnamefont  [1]{#1}%
\providecommand \bibfnamefont [1]{#1}%
\providecommand \citenamefont [1]{#1}%
\providecommand \href@noop [0]{\@secondoftwo}%
\providecommand \href [0]{\begingroup \@sanitize@url \@href}%
\providecommand \@href[1]{\@@startlink{#1}\@@href}%
\providecommand \@@href[1]{\endgroup#1\@@endlink}%
\providecommand \@sanitize@url [0]{\catcode `\\12\catcode `\$12\catcode
  `\&12\catcode `\#12\catcode `\^12\catcode `\_12\catcode `\%12\relax}%
\providecommand \@@startlink[1]{}%
\providecommand \@@endlink[0]{}%
\providecommand \url  [0]{\begingroup\@sanitize@url \@url }%
\providecommand \@url [1]{\endgroup\@href {#1}{\urlprefix }}%
\providecommand \urlprefix  [0]{URL }%
\providecommand \Eprint [0]{\href }%
\providecommand \doibase [0]{http://dx.doi.org/}%
\providecommand \selectlanguage [0]{\@gobble}%
\providecommand \bibinfo  [0]{\@secondoftwo}%
\providecommand \bibfield  [0]{\@secondoftwo}%
\providecommand \translation [1]{[#1]}%
\providecommand \BibitemOpen [0]{}%
\providecommand \bibitemStop [0]{}%
\providecommand \bibitemNoStop [0]{.\EOS\space}%
\providecommand \EOS [0]{\spacefactor3000\relax}%
\providecommand \BibitemShut  [1]{\csname bibitem#1\endcsname}%
\let\auto@bib@innerbib\@empty
\bibitem [{\citenamefont {Nortershauser}\ \emph {et~al.}(2009)\citenamefont
  {Nortershauser} \emph {et~al.}}]{Nortershauser:2008vp}%
  \BibitemOpen
  \bibfield  {author} {\bibinfo {author} {\bibfnamefont {W.}~\bibnamefont
  {Nortershauser}} \emph {et~al.},\ }\href {\doibase
  10.1103/PhysRevLett.102.062503} {\bibfield  {journal} {\bibinfo  {journal}
  {Phys. Rev. Lett.}\ }\textbf {\bibinfo {volume} {102}},\ \bibinfo {pages}
  {062503} (\bibinfo {year} {2009})},\ \Eprint {http://arxiv.org/abs/0809.2607}
  {arXiv:0809.2607 [nucl-ex]} \BibitemShut {NoStop}%
\bibitem [{\citenamefont {Geithner}\ \emph {et~al.}(2008)\citenamefont
  {Geithner} \emph {et~al.}}]{Geithner:2008zz}%
  \BibitemOpen
  \bibfield  {author} {\bibinfo {author} {\bibfnamefont {W.}~\bibnamefont
  {Geithner}} \emph {et~al.},\ }\href {\doibase 10.1103/PhysRevLett.101.252502}
  {\bibfield  {journal} {\bibinfo  {journal} {Phys. Rev. Lett.}\ }\textbf
  {\bibinfo {volume} {101}},\ \bibinfo {pages} {252502} (\bibinfo {year}
  {2008})}\BibitemShut {NoStop}%
\bibitem [{\citenamefont {Brown}(2017)}]{Brown:2017xxo}%
  \BibitemOpen
  \bibfield  {author} {\bibinfo {author} {\bibfnamefont {B.~A.}\ \bibnamefont
  {Brown}},\ }\href {\doibase 10.1103/PhysRevLett.119.122502} {\bibfield
  {journal} {\bibinfo  {journal} {Phys. Rev. Lett.}\ }\textbf {\bibinfo
  {volume} {119}},\ \bibinfo {pages} {122502} (\bibinfo {year}
  {2017})}\BibitemShut {NoStop}%
\bibitem [{\citenamefont {Yang}\ and\ \citenamefont
  {Piekarewicz}(2018)}]{Yang:2017vih}%
  \BibitemOpen
  \bibfield  {author} {\bibinfo {author} {\bibfnamefont {J.}~\bibnamefont
  {Yang}}\ and\ \bibinfo {author} {\bibfnamefont {J.}~\bibnamefont
  {Piekarewicz}},\ }\href {\doibase 10.1103/PhysRevC.97.014314} {\bibfield
  {journal} {\bibinfo  {journal} {Phys. Rev. C}\ }\textbf {\bibinfo {volume}
  {97}},\ \bibinfo {pages} {014314} (\bibinfo {year} {2018})},\ \Eprint
  {http://arxiv.org/abs/1709.10182} {arXiv:1709.10182 [nucl-th]} \BibitemShut
  {NoStop}%
\bibitem [{\citenamefont {Garcia~Ruiz}\ \emph {et~al.}(2016)\citenamefont
  {Garcia~Ruiz} \emph {et~al.}}]{GarciaRuiz:2016ohj}%
  \BibitemOpen
  \bibfield  {author} {\bibinfo {author} {\bibfnamefont {R.~F.}\ \bibnamefont
  {Garcia~Ruiz}} \emph {et~al.},\ }\href {\doibase 10.1038/nphys3645}
  {\bibfield  {journal} {\bibinfo  {journal} {Nature Phys.}\ }\textbf {\bibinfo
  {volume} {12}},\ \bibinfo {pages} {594} (\bibinfo {year} {2016})},\ \Eprint
  {http://arxiv.org/abs/1602.07906} {arXiv:1602.07906 [nucl-ex]} \BibitemShut
  {NoStop}%
\bibitem [{\citenamefont {Miller}\ \emph {et~al.}(2019)\citenamefont {Miller},
  \citenamefont {Minamisono}, \citenamefont {Klose}, \citenamefont {Garand},
  \citenamefont {Kujawa}, \citenamefont {Lantis}, \citenamefont {Liu},
  \citenamefont {Maa\ss}, \citenamefont {Mantica}, \citenamefont {Nazarewicz},
  \citenamefont {N$\mathrm{\ddot{o}}$rtersh$\mathrm{\ddot{a}}$user},
  \citenamefont {Pineda}, \citenamefont {Reinhard}, \citenamefont {Rossi},
  \citenamefont {Sommer}, \citenamefont {Sumithrarachchi}, \citenamefont
  {Teigelh$\mathrm{\ddot{o}}$fer},\ and\ \citenamefont {Watkins}}]{Miller2019}%
  \BibitemOpen
  \bibfield  {author} {\bibinfo {author} {\bibfnamefont {A.~J.}\ \bibnamefont
  {Miller}}, \bibinfo {author} {\bibfnamefont {K.}~\bibnamefont {Minamisono}},
  \bibinfo {author} {\bibfnamefont {A.}~\bibnamefont {Klose}}, \bibinfo
  {author} {\bibfnamefont {D.}~\bibnamefont {Garand}}, \bibinfo {author}
  {\bibfnamefont {C.}~\bibnamefont {Kujawa}}, \bibinfo {author} {\bibfnamefont
  {J.~D.}\ \bibnamefont {Lantis}}, \bibinfo {author} {\bibfnamefont
  {Y.}~\bibnamefont {Liu}}, \bibinfo {author} {\bibfnamefont {B.}~\bibnamefont
  {Maa\ss}}, \bibinfo {author} {\bibfnamefont {P.~F.}\ \bibnamefont {Mantica}},
  \bibinfo {author} {\bibfnamefont {W.}~\bibnamefont {Nazarewicz}}, \bibinfo
  {author} {\bibfnamefont {W.}~\bibnamefont
  {N$\mathrm{\ddot{o}}$rtersh$\mathrm{\ddot{a}}$user}}, \bibinfo {author}
  {\bibfnamefont {S.~V.}\ \bibnamefont {Pineda}}, \bibinfo {author}
  {\bibfnamefont {P.-G.}\ \bibnamefont {Reinhard}}, \bibinfo {author}
  {\bibfnamefont {D.~M.}\ \bibnamefont {Rossi}}, \bibinfo {author}
  {\bibfnamefont {F.}~\bibnamefont {Sommer}}, \bibinfo {author} {\bibfnamefont
  {C.}~\bibnamefont {Sumithrarachchi}}, \bibinfo {author} {\bibfnamefont
  {A.}~\bibnamefont {Teigelh$\mathrm{\ddot{o}}$fer}}, \ and\ \bibinfo {author}
  {\bibfnamefont {J.}~\bibnamefont {Watkins}},\ }\href {\doibase
  10.1038/s41567-019-0416-9} {\bibfield  {journal} {\bibinfo  {journal} {Nature
  Phys.}\ }\textbf {\bibinfo {volume} {15}},\ \bibinfo {pages} {432} (\bibinfo
  {year} {2019})}\BibitemShut {NoStop}%
\bibitem [{\citenamefont {An}\ \emph {et~al.}(2022{\natexlab{a}})\citenamefont
  {An}, \citenamefont {Jiang}, \citenamefont {Cao},\ and\ \citenamefont
  {Zhang}}]{An:2021yoj}%
  \BibitemOpen
  \bibfield  {author} {\bibinfo {author} {\bibfnamefont {R.}~\bibnamefont
  {An}}, \bibinfo {author} {\bibfnamefont {X.}~\bibnamefont {Jiang}}, \bibinfo
  {author} {\bibfnamefont {L.-G.}\ \bibnamefont {Cao}}, \ and\ \bibinfo
  {author} {\bibfnamefont {F.-S.}\ \bibnamefont {Zhang}},\ }\href {\doibase
  10.1103/PhysRevC.105.014325} {\bibfield  {journal} {\bibinfo  {journal}
  {Phys. Rev. C}\ }\textbf {\bibinfo {volume} {105}},\ \bibinfo {pages}
  {014325} (\bibinfo {year} {2022}{\natexlab{a}})},\ \Eprint
  {http://arxiv.org/abs/2108.00278} {arXiv:2108.00278 [nucl-th]} \BibitemShut
  {NoStop}%
\bibitem [{\citenamefont {Marsh}\ \emph {et~al.}(2018)\citenamefont {Marsh}
  \emph {et~al.}}]{Marsh:2018wxs}%
  \BibitemOpen
  \bibfield  {author} {\bibinfo {author} {\bibfnamefont {B.~A.}\ \bibnamefont
  {Marsh}} \emph {et~al.},\ }\href {\doibase 10.1038/s41567-018-0292-8}
  {\bibfield  {journal} {\bibinfo  {journal} {Nature Phys.}\ }\textbf {\bibinfo
  {volume} {14}},\ \bibinfo {pages} {1163} (\bibinfo {year}
  {2018})}\BibitemShut {NoStop}%
\bibitem [{\citenamefont {Barzakh}\ \emph {et~al.}(2021)\citenamefont {Barzakh}
  \emph {et~al.}}]{Barzakh:2021gfl}%
  \BibitemOpen
  \bibfield  {author} {\bibinfo {author} {\bibfnamefont {A.}~\bibnamefont
  {Barzakh}} \emph {et~al.},\ }\href {\doibase 10.1103/PhysRevLett.127.192501}
  {\bibfield  {journal} {\bibinfo  {journal} {Phys. Rev. Lett.}\ }\textbf
  {\bibinfo {volume} {127}},\ \bibinfo {pages} {192501} (\bibinfo {year}
  {2021})}\BibitemShut {NoStop}%
\bibitem [{\citenamefont {Angeli}\ and\ \citenamefont
  {Marinova}(2015)}]{Angeli:2015wia}%
  \BibitemOpen
  \bibfield  {author} {\bibinfo {author} {\bibfnamefont {I.}~\bibnamefont
  {Angeli}}\ and\ \bibinfo {author} {\bibfnamefont {K.~P.}\ \bibnamefont
  {Marinova}},\ }\href {\doibase 10.1088/0954-3899/42/5/055108} {\bibfield
  {journal} {\bibinfo  {journal} {J. Phys. G}\ }\textbf {\bibinfo {volume}
  {42}},\ \bibinfo {pages} {055108} (\bibinfo {year} {2015})}\BibitemShut
  {NoStop}%
\bibitem [{\citenamefont {Angeli}\ and\ \citenamefont
  {Marinova}(2013)}]{Angeli:2013epw}%
  \BibitemOpen
  \bibfield  {author} {\bibinfo {author} {\bibfnamefont {I.}~\bibnamefont
  {Angeli}}\ and\ \bibinfo {author} {\bibfnamefont {K.~P.}\ \bibnamefont
  {Marinova}},\ }\href {\doibase 10.1016/j.adt.2011.12.006} {\bibfield
  {journal} {\bibinfo  {journal} {At. Data Nucl. Data Tables}\ }\textbf
  {\bibinfo {volume} {99}},\ \bibinfo {pages} {69} (\bibinfo {year}
  {2013})}\BibitemShut {NoStop}%
\bibitem [{\citenamefont {Li}\ \emph {et~al.}(2021)\citenamefont {Li},
  \citenamefont {Luo},\ and\ \citenamefont {Wang}}]{Li:2021fmk}%
  \BibitemOpen
  \bibfield  {author} {\bibinfo {author} {\bibfnamefont {T.}~\bibnamefont
  {Li}}, \bibinfo {author} {\bibfnamefont {Y.}~\bibnamefont {Luo}}, \ and\
  \bibinfo {author} {\bibfnamefont {N.}~\bibnamefont {Wang}},\ }\href {\doibase
  10.1016/j.adt.2021.101440} {\bibfield  {journal} {\bibinfo  {journal} {At.
  Data Nucl. Data Tables}\ }\textbf {\bibinfo {volume} {140}},\ \bibinfo
  {pages} {101440} (\bibinfo {year} {2021})}\BibitemShut {NoStop}%
\bibitem [{\citenamefont {Pineda}\ \emph {et~al.}(2021)\citenamefont {Pineda}
  \emph {et~al.}}]{Pineda:2021shy}%
  \BibitemOpen
  \bibfield  {author} {\bibinfo {author} {\bibfnamefont {S.~V.}\ \bibnamefont
  {Pineda}} \emph {et~al.},\ }\href {\doibase 10.1103/PhysRevLett.127.182503}
  {\bibfield  {journal} {\bibinfo  {journal} {Phys. Rev. Lett.}\ }\textbf
  {\bibinfo {volume} {127}},\ \bibinfo {pages} {182503} (\bibinfo {year}
  {2021})},\ \Eprint {http://arxiv.org/abs/2106.10378} {arXiv:2106.10378
  [nucl-ex]} \BibitemShut {NoStop}%
\bibitem [{\citenamefont {Malbrunot-Ettenauer}\ \emph
  {et~al.}(2022)\citenamefont {Malbrunot-Ettenauer} \emph
  {et~al.}}]{Malbrunot-Ettenauer:2021fnr}%
  \BibitemOpen
  \bibfield  {author} {\bibinfo {author} {\bibfnamefont {S.}~\bibnamefont
  {Malbrunot-Ettenauer}} \emph {et~al.},\ }\href {\doibase
  10.1103/PhysRevLett.128.022502} {\bibfield  {journal} {\bibinfo  {journal}
  {Phys. Rev. Lett.}\ }\textbf {\bibinfo {volume} {128}},\ \bibinfo {pages}
  {022502} (\bibinfo {year} {2022})},\ \Eprint
  {http://arxiv.org/abs/2112.03382} {arXiv:2112.03382 [nucl-ex]} \BibitemShut
  {NoStop}%
\bibitem [{\citenamefont {Reponen}\ \emph {et~al.}(2021)\citenamefont {Reponen}
  \emph {et~al.}}]{Reponen:2021rwy}%
  \BibitemOpen
  \bibfield  {author} {\bibinfo {author} {\bibfnamefont {M.}~\bibnamefont
  {Reponen}} \emph {et~al.},\ }\href {\doibase 10.1038/s41467-021-24888-x}
  {\bibfield  {journal} {\bibinfo  {journal} {Nature Commun.}\ }\textbf
  {\bibinfo {volume} {12}},\ \bibinfo {pages} {4596} (\bibinfo {year}
  {2021})}\BibitemShut {NoStop}%
\bibitem [{\citenamefont {Weizsacker}(1935)}]{Weizsacker:1935bkz}%
  \BibitemOpen
  \bibfield  {author} {\bibinfo {author} {\bibfnamefont {C.~F.~V.}\
  \bibnamefont {Weizsacker}},\ }\href {\doibase 10.1007/BF01337700} {\bibfield
  {journal} {\bibinfo  {journal} {Z. Phys.}\ }\textbf {\bibinfo {volume}
  {96}},\ \bibinfo {pages} {431} (\bibinfo {year} {1935})}\BibitemShut
  {NoStop}%
\bibitem [{\citenamefont {Brown}\ \emph {et~al.}(1984)\citenamefont {Brown},
  \citenamefont {Bronk},\ and\ \citenamefont {Hodgson}}]{Brown:1984zz}%
  \BibitemOpen
  \bibfield  {author} {\bibinfo {author} {\bibfnamefont {B.~A.}\ \bibnamefont
  {Brown}}, \bibinfo {author} {\bibfnamefont {C.~R.}\ \bibnamefont {Bronk}}, \
  and\ \bibinfo {author} {\bibfnamefont {P.~E.}\ \bibnamefont {Hodgson}},\
  }\href {\doibase 10.1088/0305-4616/10/12/008} {\bibfield  {journal} {\bibinfo
   {journal} {J. Phys. G}\ }\textbf {\bibinfo {volume} {10}},\ \bibinfo {pages}
  {1683} (\bibinfo {year} {1984})}\BibitemShut {NoStop}%
\bibitem [{\citenamefont {Nerlo-Pomorska}\ and\ \citenamefont
  {Pomorski}(1994)}]{Nerlo-Pomorska:1994dhg}%
  \BibitemOpen
  \bibfield  {author} {\bibinfo {author} {\bibfnamefont {B.}~\bibnamefont
  {Nerlo-Pomorska}}\ and\ \bibinfo {author} {\bibfnamefont {K.}~\bibnamefont
  {Pomorski}},\ }\href {\doibase 10.1007/BF01291913} {\bibfield  {journal}
  {\bibinfo  {journal} {Z. Phys. A}\ }\textbf {\bibinfo {volume} {348}},\
  \bibinfo {pages} {169} (\bibinfo {year} {1994})},\ \Eprint
  {http://arxiv.org/abs/nucl-th/9401015} {arXiv:nucl-th/9401015} \BibitemShut
  {NoStop}%
\bibitem [{\citenamefont {Zhang}\ \emph {et~al.}(2002)\citenamefont {Zhang},
  \citenamefont {Meng}, \citenamefont {Zhou},\ and\ \citenamefont
  {Zeng}}]{Zhang:2001nt}%
  \BibitemOpen
  \bibfield  {author} {\bibinfo {author} {\bibfnamefont {S.~Q.}\ \bibnamefont
  {Zhang}}, \bibinfo {author} {\bibfnamefont {J.}~\bibnamefont {Meng}},
  \bibinfo {author} {\bibfnamefont {S.~G.}\ \bibnamefont {Zhou}}, \ and\
  \bibinfo {author} {\bibfnamefont {J.~Y.}\ \bibnamefont {Zeng}},\ }\href
  {\doibase 10.1007/s10050-002-8757-6} {\bibfield  {journal} {\bibinfo
  {journal} {Eur. Phys. J. A}\ }\textbf {\bibinfo {volume} {13}},\ \bibinfo
  {pages} {285} (\bibinfo {year} {2002})},\ \Eprint
  {http://arxiv.org/abs/nucl-th/0107040} {arXiv:nucl-th/0107040} \BibitemShut
  {NoStop}%
\bibitem [{\citenamefont {Wang}\ and\ \citenamefont {Li}(2013)}]{Wang:2013zia}%
  \BibitemOpen
  \bibfield  {author} {\bibinfo {author} {\bibfnamefont {N.}~\bibnamefont
  {Wang}}\ and\ \bibinfo {author} {\bibfnamefont {T.}~\bibnamefont {Li}},\
  }\href {\doibase 10.1103/PhysRevC.88.011301} {\bibfield  {journal} {\bibinfo
  {journal} {Phys. Rev. C}\ }\textbf {\bibinfo {volume} {88}},\ \bibinfo
  {pages} {011301} (\bibinfo {year} {2013})},\ \Eprint
  {http://arxiv.org/abs/1307.2315} {arXiv:1307.2315 [nucl-th]} \BibitemShut
  {NoStop}%
\bibitem [{\citenamefont {Sheng}\ \emph {et~al.}(2015)\citenamefont {Sheng},
  \citenamefont {Fan}, \citenamefont {Qian},\ and\ \citenamefont
  {Hu}}]{Sheng:2015poa}%
  \BibitemOpen
  \bibfield  {author} {\bibinfo {author} {\bibfnamefont {Z.}~\bibnamefont
  {Sheng}}, \bibinfo {author} {\bibfnamefont {G.}~\bibnamefont {Fan}}, \bibinfo
  {author} {\bibfnamefont {J.}~\bibnamefont {Qian}}, \ and\ \bibinfo {author}
  {\bibfnamefont {J.}~\bibnamefont {Hu}},\ }\href {\doibase
  10.1140/epja/i2015-15040-1} {\bibfield  {journal} {\bibinfo  {journal} {Eur.
  Phys. J. A}\ }\textbf {\bibinfo {volume} {51}},\ \bibinfo {pages} {40}
  (\bibinfo {year} {2015})}\BibitemShut {NoStop}%
\bibitem [{\citenamefont {Peng}\ \emph {et~al.}(2014)\citenamefont {Peng},
  \citenamefont {Lu}, \citenamefont {Sun},\ and\ \citenamefont
  {Zhao}}]{Peng:2014jia}%
  \BibitemOpen
  \bibfield  {author} {\bibinfo {author} {\bibfnamefont {J.~J.}\ \bibnamefont
  {Peng}}, \bibinfo {author} {\bibfnamefont {Y.}~\bibnamefont {Lu}}, \bibinfo
  {author} {\bibfnamefont {B.~H.}\ \bibnamefont {Sun}}, \ and\ \bibinfo
  {author} {\bibfnamefont {Y.~M.}\ \bibnamefont {Zhao}},\ }\href {\doibase
  10.1103/PhysRevC.90.054318} {\bibfield  {journal} {\bibinfo  {journal} {Phys.
  Rev. C}\ }\textbf {\bibinfo {volume} {90}},\ \bibinfo {pages} {054318}
  (\bibinfo {year} {2014})},\ \bibinfo {note} {[Erratum: Phys.Rev.C 91, 019902
  (2015)]},\ \Eprint {http://arxiv.org/abs/1408.6954} {arXiv:1408.6954
  [nucl-th]} \BibitemShut {NoStop}%
\bibitem [{\citenamefont {Bao}\ \emph {et~al.}(2016)\citenamefont {Bao},
  \citenamefont {Lu}, \citenamefont {Zhao},\ and\ \citenamefont
  {Arima}}]{Bao:2016suw}%
  \BibitemOpen
  \bibfield  {author} {\bibinfo {author} {\bibfnamefont {M.}~\bibnamefont
  {Bao}}, \bibinfo {author} {\bibfnamefont {Y.}~\bibnamefont {Lu}}, \bibinfo
  {author} {\bibfnamefont {Y.~M.}\ \bibnamefont {Zhao}}, \ and\ \bibinfo
  {author} {\bibfnamefont {A.}~\bibnamefont {Arima}},\ }\href {\doibase
  10.1103/PhysRevC.94.064315} {\bibfield  {journal} {\bibinfo  {journal} {Phys.
  Rev. C}\ }\textbf {\bibinfo {volume} {94}},\ \bibinfo {pages} {064315}
  (\bibinfo {year} {2016})}\BibitemShut {NoStop}%
\bibitem [{\citenamefont {Sun}\ \emph {et~al.}(2017)\citenamefont {Sun},
  \citenamefont {Liu},\ and\ \citenamefont {Wang}}]{Sun:2016nec}%
  \BibitemOpen
  \bibfield  {author} {\bibinfo {author} {\bibfnamefont {B.-H.}\ \bibnamefont
  {Sun}}, \bibinfo {author} {\bibfnamefont {C.-Y.}\ \bibnamefont {Liu}}, \ and\
  \bibinfo {author} {\bibfnamefont {H.-X.}\ \bibnamefont {Wang}},\ }\href
  {\doibase 10.1103/PhysRevC.95.014307} {\bibfield  {journal} {\bibinfo
  {journal} {Phys. Rev. C}\ }\textbf {\bibinfo {volume} {95}},\ \bibinfo
  {pages} {014307} (\bibinfo {year} {2017})},\ \Eprint
  {http://arxiv.org/abs/1609.03144} {arXiv:1609.03144 [nucl-th]} \BibitemShut
  {NoStop}%
\bibitem [{\citenamefont {Bao}\ \emph {et~al.}(2020)\citenamefont {Bao},
  \citenamefont {Zong}, \citenamefont {Zhao},\ and\ \citenamefont
  {Arima}}]{Bao:2020ffv}%
  \BibitemOpen
  \bibfield  {author} {\bibinfo {author} {\bibfnamefont {M.}~\bibnamefont
  {Bao}}, \bibinfo {author} {\bibfnamefont {Y.~Y.}\ \bibnamefont {Zong}},
  \bibinfo {author} {\bibfnamefont {Y.~M.}\ \bibnamefont {Zhao}}, \ and\
  \bibinfo {author} {\bibfnamefont {A.}~\bibnamefont {Arima}},\ }\href
  {\doibase 10.1103/PhysRevC.102.014306} {\bibfield  {journal} {\bibinfo
  {journal} {Phys. Rev. C}\ }\textbf {\bibinfo {volume} {102}},\ \bibinfo
  {pages} {014306} (\bibinfo {year} {2020})}\BibitemShut {NoStop}%
\bibitem [{\citenamefont {Ma}\ \emph {et~al.}(2021)\citenamefont {Ma},
  \citenamefont {Zong}, \citenamefont {Zhao},\ and\ \citenamefont
  {Arima}}]{Ma:2021jzu}%
  \BibitemOpen
  \bibfield  {author} {\bibinfo {author} {\bibfnamefont {C.}~\bibnamefont
  {Ma}}, \bibinfo {author} {\bibfnamefont {Y.~Y.}\ \bibnamefont {Zong}},
  \bibinfo {author} {\bibfnamefont {Y.~M.}\ \bibnamefont {Zhao}}, \ and\
  \bibinfo {author} {\bibfnamefont {A.}~\bibnamefont {Arima}},\ }\href
  {\doibase 10.1103/PhysRevC.104.014303} {\bibfield  {journal} {\bibinfo
  {journal} {Phys. Rev. C}\ }\textbf {\bibinfo {volume} {104}},\ \bibinfo
  {pages} {014303} (\bibinfo {year} {2021})}\BibitemShut {NoStop}%
\bibitem [{\citenamefont {Geng}\ \emph {et~al.}(2005)\citenamefont {Geng},
  \citenamefont {Toki},\ and\ \citenamefont {Meng}}]{Geng:2005yu}%
  \BibitemOpen
  \bibfield  {author} {\bibinfo {author} {\bibfnamefont {L.-S.}\ \bibnamefont
  {Geng}}, \bibinfo {author} {\bibfnamefont {H.}~\bibnamefont {Toki}}, \ and\
  \bibinfo {author} {\bibfnamefont {J.}~\bibnamefont {Meng}},\ }\href {\doibase
  10.1143/PTP.113.785} {\bibfield  {journal} {\bibinfo  {journal} {Prog. Theor.
  Phys.}\ }\textbf {\bibinfo {volume} {113}},\ \bibinfo {pages} {785} (\bibinfo
  {year} {2005})},\ \Eprint {http://arxiv.org/abs/nucl-th/0503086}
  {arXiv:nucl-th/0503086} \BibitemShut {NoStop}%
\bibitem [{\citenamefont {Goriely}\ \emph {et~al.}(2016)\citenamefont
  {Goriely}, \citenamefont {Chamel},\ and\ \citenamefont
  {Pearson}}]{Goriely:2016sdz}%
  \BibitemOpen
  \bibfield  {author} {\bibinfo {author} {\bibfnamefont {S.}~\bibnamefont
  {Goriely}}, \bibinfo {author} {\bibfnamefont {N.}~\bibnamefont {Chamel}}, \
  and\ \bibinfo {author} {\bibfnamefont {J.~M.}\ \bibnamefont {Pearson}},\
  }\href {\doibase 10.1103/PhysRevC.93.034337} {\bibfield  {journal} {\bibinfo
  {journal} {Phys. Rev. C}\ }\textbf {\bibinfo {volume} {93}},\ \bibinfo
  {pages} {034337} (\bibinfo {year} {2016})}\BibitemShut {NoStop}%
\bibitem [{\citenamefont {Pe\~na Arteaga}\ \emph {et~al.}(2016)\citenamefont
  {Pe\~na Arteaga}, \citenamefont {Goriely},\ and\ \citenamefont
  {Chamel}}]{Pena-Arteaga:2016clz}%
  \BibitemOpen
  \bibfield  {author} {\bibinfo {author} {\bibfnamefont {D.}~\bibnamefont
  {Pe\~na Arteaga}}, \bibinfo {author} {\bibfnamefont {S.}~\bibnamefont
  {Goriely}}, \ and\ \bibinfo {author} {\bibfnamefont {N.}~\bibnamefont
  {Chamel}},\ }\href {\doibase 10.1140/epja/i2016-16320-x} {\bibfield
  {journal} {\bibinfo  {journal} {Eur. Phys. J. A}\ }\textbf {\bibinfo {volume}
  {52}},\ \bibinfo {pages} {320} (\bibinfo {year} {2016})}\BibitemShut
  {NoStop}%
\bibitem [{\citenamefont {Sarriguren}(2019)}]{Sarriguren:2019jfb}%
  \BibitemOpen
  \bibfield  {author} {\bibinfo {author} {\bibfnamefont {P.}~\bibnamefont
  {Sarriguren}},\ }\href {\doibase 10.1103/PhysRevC.100.054306} {\bibfield
  {journal} {\bibinfo  {journal} {Phys. Rev. C}\ }\textbf {\bibinfo {volume}
  {100}},\ \bibinfo {pages} {054306} (\bibinfo {year} {2019})},\ \Eprint
  {http://arxiv.org/abs/1911.04313} {arXiv:1911.04313 [nucl-th]} \BibitemShut
  {NoStop}%
\bibitem [{\citenamefont {An}\ \emph {et~al.}(2020)\citenamefont {An},
  \citenamefont {Geng},\ and\ \citenamefont {Zhang}}]{An:2020qgp}%
  \BibitemOpen
  \bibfield  {author} {\bibinfo {author} {\bibfnamefont {R.}~\bibnamefont
  {An}}, \bibinfo {author} {\bibfnamefont {L.-S.}\ \bibnamefont {Geng}}, \ and\
  \bibinfo {author} {\bibfnamefont {S.-S.}\ \bibnamefont {Zhang}},\ }\href
  {\doibase 10.1103/PhysRevC.102.024307} {\bibfield  {journal} {\bibinfo
  {journal} {Phys. Rev. C}\ }\textbf {\bibinfo {volume} {102}},\ \bibinfo
  {pages} {024307} (\bibinfo {year} {2020})},\ \Eprint
  {http://arxiv.org/abs/2005.00141} {arXiv:2005.00141 [nucl-th]} \BibitemShut
  {NoStop}%
\bibitem [{\citenamefont {An}\ \emph {et~al.}(2022{\natexlab{b}})\citenamefont
  {An}, \citenamefont {Zhang}, \citenamefont {Geng},\ and\ \citenamefont
  {Zhang}}]{An:2021rlw}%
  \BibitemOpen
  \bibfield  {author} {\bibinfo {author} {\bibfnamefont {R.}~\bibnamefont
  {An}}, \bibinfo {author} {\bibfnamefont {S.-S.}\ \bibnamefont {Zhang}},
  \bibinfo {author} {\bibfnamefont {L.-S.}\ \bibnamefont {Geng}}, \ and\
  \bibinfo {author} {\bibfnamefont {F.-S.}\ \bibnamefont {Zhang}},\ }\href
  {\doibase 10.1088/1674-1137/ac4b5c} {\bibfield  {journal} {\bibinfo
  {journal} {Chin. Phys. C}\ }\textbf {\bibinfo {volume} {46}},\ \bibinfo
  {pages} {054101} (\bibinfo {year} {2022}{\natexlab{b}})},\ \Eprint
  {http://arxiv.org/abs/2106.02279} {arXiv:2106.02279 [nucl-th]} \BibitemShut
  {NoStop}%
\bibitem [{\citenamefont {Zhang}\ \emph {et~al.}(2022)\citenamefont {Zhang}
  \emph {et~al.}}]{DRHBcMassTable:2022uhi}%
  \BibitemOpen
  \bibfield  {author} {\bibinfo {author} {\bibfnamefont {K.}~\bibnamefont
  {Zhang}} \emph {et~al.} (\bibinfo {collaboration} {DRHBc Mass Table}),\
  }\href {\doibase 10.1016/j.adt.2022.101488} {\bibfield  {journal} {\bibinfo
  {journal} {At. Data Nucl. Data Tables}\ }\textbf {\bibinfo {volume} {144}},\
  \bibinfo {pages} {101488} (\bibinfo {year} {2022})},\ \Eprint
  {http://arxiv.org/abs/2201.03216} {arXiv:2201.03216 [nucl-th]} \BibitemShut
  {NoStop}%
\bibitem [{\citenamefont {Forssen}\ \emph {et~al.}(2009)\citenamefont
  {Forssen}, \citenamefont {Caurier},\ and\ \citenamefont
  {Navratil}}]{Forssen:2009vu}%
  \BibitemOpen
  \bibfield  {author} {\bibinfo {author} {\bibfnamefont {C.}~\bibnamefont
  {Forssen}}, \bibinfo {author} {\bibfnamefont {E.}~\bibnamefont {Caurier}}, \
  and\ \bibinfo {author} {\bibfnamefont {P.}~\bibnamefont {Navratil}},\ }\href
  {\doibase 10.1103/PhysRevC.79.021303} {\bibfield  {journal} {\bibinfo
  {journal} {Phys. Rev. C}\ }\textbf {\bibinfo {volume} {79}},\ \bibinfo
  {pages} {021303} (\bibinfo {year} {2009})},\ \Eprint
  {http://arxiv.org/abs/0901.0453} {arXiv:0901.0453 [nucl-th]} \BibitemShut
  {NoStop}%
\bibitem [{\citenamefont {Reinhard}\ and\ \citenamefont
  {Nazarewicz}(2017)}]{Reinhard:2017ugx}%
  \BibitemOpen
  \bibfield  {author} {\bibinfo {author} {\bibfnamefont {P.~G.}\ \bibnamefont
  {Reinhard}}\ and\ \bibinfo {author} {\bibfnamefont {W.}~\bibnamefont
  {Nazarewicz}},\ }\href {\doibase 10.1103/PhysRevC.95.064328} {\bibfield
  {journal} {\bibinfo  {journal} {Phys. Rev. C}\ }\textbf {\bibinfo {volume}
  {95}},\ \bibinfo {pages} {064328} (\bibinfo {year} {2017})},\ \Eprint
  {http://arxiv.org/abs/1704.07430} {arXiv:1704.07430 [nucl-th]} \BibitemShut
  {NoStop}%
\bibitem [{\citenamefont {Carleo}\ \emph {et~al.}(2019)\citenamefont {Carleo},
  \citenamefont {Cirac}, \citenamefont {Cranmer}, \citenamefont {Daudet},
  \citenamefont {Schuld}, \citenamefont {Tishby}, \citenamefont
  {Vogt-Maranto},\ and\ \citenamefont {Zdeborov\'a}}]{Carleo:2019ptp}%
  \BibitemOpen
  \bibfield  {author} {\bibinfo {author} {\bibfnamefont {G.}~\bibnamefont
  {Carleo}}, \bibinfo {author} {\bibfnamefont {I.}~\bibnamefont {Cirac}},
  \bibinfo {author} {\bibfnamefont {K.}~\bibnamefont {Cranmer}}, \bibinfo
  {author} {\bibfnamefont {L.}~\bibnamefont {Daudet}}, \bibinfo {author}
  {\bibfnamefont {M.}~\bibnamefont {Schuld}}, \bibinfo {author} {\bibfnamefont
  {N.}~\bibnamefont {Tishby}}, \bibinfo {author} {\bibfnamefont
  {L.}~\bibnamefont {Vogt-Maranto}}, \ and\ \bibinfo {author} {\bibfnamefont
  {L.}~\bibnamefont {Zdeborov\'a}},\ }\href {\doibase
  10.1103/RevModPhys.91.045002} {\bibfield  {journal} {\bibinfo  {journal}
  {Rev. Mod. Phys.}\ }\textbf {\bibinfo {volume} {91}},\ \bibinfo {pages}
  {045002} (\bibinfo {year} {2019})},\ \Eprint
  {http://arxiv.org/abs/1903.10563} {arXiv:1903.10563 [physics.comp-ph]}
  \BibitemShut {NoStop}%
\bibitem [{\citenamefont {Bourilkov}(2020)}]{Bourilkov:2019yoi}%
  \BibitemOpen
  \bibfield  {author} {\bibinfo {author} {\bibfnamefont {D.}~\bibnamefont
  {Bourilkov}},\ }\href {\doibase 10.1142/S0217751X19300199} {\bibfield
  {journal} {\bibinfo  {journal} {Int. J. Mod. Phys. A}\ }\textbf {\bibinfo
  {volume} {34}},\ \bibinfo {pages} {1930019} (\bibinfo {year} {2020})},\
  \Eprint {http://arxiv.org/abs/1912.08245} {arXiv:1912.08245
  [physics.data-an]} \BibitemShut {NoStop}%
\bibitem [{\citenamefont {Bedolla-Montiel}\ \emph {et~al.}(2021)\citenamefont
  {Bedolla-Montiel}, \citenamefont {Padierna},\ and\ \citenamefont {Casta\~neda
  Priego}}]{Bedolla-Montiel:2020rio}%
  \BibitemOpen
  \bibfield  {author} {\bibinfo {author} {\bibfnamefont {E.~A.}\ \bibnamefont
  {Bedolla-Montiel}}, \bibinfo {author} {\bibfnamefont {L.~C.}\ \bibnamefont
  {Padierna}}, \ and\ \bibinfo {author} {\bibfnamefont {R.}~\bibnamefont
  {Casta\~neda Priego}},\ }\href {\doibase 10.1088/1361-648X/abb895} {\bibfield
   {journal} {\bibinfo  {journal} {J. Phys. Condens. Matter}\ }\textbf
  {\bibinfo {volume} {33}},\ \bibinfo {pages} {053001} (\bibinfo {year}
  {2021})},\ \Eprint {http://arxiv.org/abs/2005.14228} {arXiv:2005.14228
  [physics.comp-ph]} \BibitemShut {NoStop}%
\bibitem [{\citenamefont {Bedaque}\ \emph {et~al.}(2021)\citenamefont {Bedaque}
  \emph {et~al.}}]{Bedaque:2021bja}%
  \BibitemOpen
  \bibfield  {author} {\bibinfo {author} {\bibfnamefont {P.}~\bibnamefont
  {Bedaque}} \emph {et~al.},\ }\href {\doibase 10.1140/epja/s10050-020-00290-x}
  {\bibfield  {journal} {\bibinfo  {journal} {Eur. Phys. J. A}\ }\textbf
  {\bibinfo {volume} {57}},\ \bibinfo {pages} {100} (\bibinfo {year}
  {2021})}\BibitemShut {NoStop}%
\bibitem [{\citenamefont {Boehnlein}\ \emph {et~al.}(2021)\citenamefont
  {Boehnlein} \emph {et~al.}}]{Boehnlein:2021eym}%
  \BibitemOpen
  \bibfield  {author} {\bibinfo {author} {\bibfnamefont {A.}~\bibnamefont
  {Boehnlein}} \emph {et~al.},\ }\href@noop {} {\  (\bibinfo {year} {2021})},\
  \Eprint {http://arxiv.org/abs/2112.02309} {arXiv:2112.02309 [nucl-th]}
  \BibitemShut {NoStop}%
\bibitem [{\citenamefont {Ma}\ \emph {et~al.}(2020)\citenamefont {Ma},
  \citenamefont {Su}, \citenamefont {Liu}, \citenamefont {Ren}, \citenamefont
  {Xu},\ and\ \citenamefont {Gao}}]{Ma:2020rdk}%
  \BibitemOpen
  \bibfield  {author} {\bibinfo {author} {\bibfnamefont {Y.}~\bibnamefont
  {Ma}}, \bibinfo {author} {\bibfnamefont {C.}~\bibnamefont {Su}}, \bibinfo
  {author} {\bibfnamefont {J.}~\bibnamefont {Liu}}, \bibinfo {author}
  {\bibfnamefont {Z.}~\bibnamefont {Ren}}, \bibinfo {author} {\bibfnamefont
  {C.}~\bibnamefont {Xu}}, \ and\ \bibinfo {author} {\bibfnamefont
  {Y.}~\bibnamefont {Gao}},\ }\href {\doibase 10.1103/PhysRevC.101.014304}
  {\bibfield  {journal} {\bibinfo  {journal} {Phys. Rev. C}\ }\textbf {\bibinfo
  {volume} {101}},\ \bibinfo {pages} {014304} (\bibinfo {year}
  {2020})}\BibitemShut {NoStop}%
\bibitem [{\citenamefont {Ma}\ and\ \citenamefont {Zhang}(2022)}]{Ma:2022yrj}%
  \BibitemOpen
  \bibfield  {author} {\bibinfo {author} {\bibfnamefont {J.-Q.}\ \bibnamefont
  {Ma}}\ and\ \bibinfo {author} {\bibfnamefont {Z.-H.}\ \bibnamefont {Zhang}},\
  }\href@noop {} {\  (\bibinfo {year} {2022})},\ \Eprint
  {http://arxiv.org/abs/2203.14027} {arXiv:2203.14027 [nucl-th]} \BibitemShut
  {NoStop}%
\bibitem [{\citenamefont {Akkoyun}\ \emph {et~al.}(2013)\citenamefont
  {Akkoyun}, \citenamefont {Bayram}, \citenamefont {Kara},\ and\ \citenamefont
  {Sinan}}]{Akkoyun:2012yf}%
  \BibitemOpen
  \bibfield  {author} {\bibinfo {author} {\bibfnamefont {S.}~\bibnamefont
  {Akkoyun}}, \bibinfo {author} {\bibfnamefont {T.}~\bibnamefont {Bayram}},
  \bibinfo {author} {\bibfnamefont {S.~O.}\ \bibnamefont {Kara}}, \ and\
  \bibinfo {author} {\bibfnamefont {A.}~\bibnamefont {Sinan}},\ }\href
  {\doibase 10.1088/0954-3899/40/5/055106} {\bibfield  {journal} {\bibinfo
  {journal} {J. Phys. G}\ }\textbf {\bibinfo {volume} {40}},\ \bibinfo {pages}
  {055106} (\bibinfo {year} {2013})},\ \Eprint {http://arxiv.org/abs/1212.6319}
  {arXiv:1212.6319 [nucl-th]} \BibitemShut {NoStop}%
\bibitem [{\citenamefont {Wu}\ \emph {et~al.}(2020)\citenamefont {Wu},
  \citenamefont {Bai}, \citenamefont {Sagawa},\ and\ \citenamefont
  {Zhang}}]{Wu:2020bao}%
  \BibitemOpen
  \bibfield  {author} {\bibinfo {author} {\bibfnamefont {D.}~\bibnamefont
  {Wu}}, \bibinfo {author} {\bibfnamefont {C.~L.}\ \bibnamefont {Bai}},
  \bibinfo {author} {\bibfnamefont {H.}~\bibnamefont {Sagawa}}, \ and\ \bibinfo
  {author} {\bibfnamefont {H.~Q.}\ \bibnamefont {Zhang}},\ }\href {\doibase
  10.1103/PhysRevC.102.054323} {\bibfield  {journal} {\bibinfo  {journal}
  {Phys. Rev. C}\ }\textbf {\bibinfo {volume} {102}},\ \bibinfo {pages}
  {054323} (\bibinfo {year} {2020})},\ \Eprint
  {http://arxiv.org/abs/2006.09677} {arXiv:2006.09677 [nucl-th]} \BibitemShut
  {NoStop}%
\bibitem [{\citenamefont {Yang}\ \emph {et~al.}(2022)\citenamefont {Yang},
  \citenamefont {Fan}, \citenamefont {Naito}, \citenamefont {Niu},
  \citenamefont {Li},\ and\ \citenamefont {Liang}}]{Yang:2022tjf}%
  \BibitemOpen
  \bibfield  {author} {\bibinfo {author} {\bibfnamefont {Z.-X.}\ \bibnamefont
  {Yang}}, \bibinfo {author} {\bibfnamefont {X.-H.}\ \bibnamefont {Fan}},
  \bibinfo {author} {\bibfnamefont {T.}~\bibnamefont {Naito}}, \bibinfo
  {author} {\bibfnamefont {Z.-M.}\ \bibnamefont {Niu}}, \bibinfo {author}
  {\bibfnamefont {Z.-P.}\ \bibnamefont {Li}}, \ and\ \bibinfo {author}
  {\bibfnamefont {H.}~\bibnamefont {Liang}},\ }\href@noop {} {\  (\bibinfo
  {year} {2022})},\ \Eprint {http://arxiv.org/abs/2205.15649} {arXiv:2205.15649
  [nucl-th]} \BibitemShut {NoStop}%
\bibitem [{\citenamefont {Utama}\ \emph {et~al.}(2016)\citenamefont {Utama},
  \citenamefont {Chen},\ and\ \citenamefont {Piekarewicz}}]{Utama:2016tcl}%
  \BibitemOpen
  \bibfield  {author} {\bibinfo {author} {\bibfnamefont {R.}~\bibnamefont
  {Utama}}, \bibinfo {author} {\bibfnamefont {W.-C.}\ \bibnamefont {Chen}}, \
  and\ \bibinfo {author} {\bibfnamefont {J.}~\bibnamefont {Piekarewicz}},\
  }\href {\doibase 10.1088/0954-3899/43/11/114002} {\bibfield  {journal}
  {\bibinfo  {journal} {J. Phys. G}\ }\textbf {\bibinfo {volume} {43}},\
  \bibinfo {pages} {114002} (\bibinfo {year} {2016})},\ \Eprint
  {http://arxiv.org/abs/1608.03020} {arXiv:1608.03020 [nucl-th]} \BibitemShut
  {NoStop}%
\bibitem [{\citenamefont {Dong}\ \emph {et~al.}(2022)\citenamefont {Dong},
  \citenamefont {An}, \citenamefont {Lu},\ and\ \citenamefont
  {Geng}}]{Dong:2021aqg}%
  \BibitemOpen
  \bibfield  {author} {\bibinfo {author} {\bibfnamefont {X.-X.}\ \bibnamefont
  {Dong}}, \bibinfo {author} {\bibfnamefont {R.}~\bibnamefont {An}}, \bibinfo
  {author} {\bibfnamefont {J.-X.}\ \bibnamefont {Lu}}, \ and\ \bibinfo {author}
  {\bibfnamefont {L.-S.}\ \bibnamefont {Geng}},\ }\href {\doibase
  10.1103/PhysRevC.105.014308} {\bibfield  {journal} {\bibinfo  {journal}
  {Phys. Rev. C}\ }\textbf {\bibinfo {volume} {105}},\ \bibinfo {pages}
  {014308} (\bibinfo {year} {2022})},\ \Eprint
  {http://arxiv.org/abs/2109.09626} {arXiv:2109.09626 [nucl-th]} \BibitemShut
  {NoStop}%
\bibitem [{\citenamefont {Day~Goodacre}\ \emph {et~al.}(2021)\citenamefont
  {Day~Goodacre} \emph {et~al.}}]{Goodacre:2020sys}%
  \BibitemOpen
  \bibfield  {author} {\bibinfo {author} {\bibfnamefont {T.}~\bibnamefont
  {Day~Goodacre}} \emph {et~al.},\ }\href {\doibase
  10.1103/PhysRevLett.126.032502} {\bibfield  {journal} {\bibinfo  {journal}
  {Phys. Rev. Lett.}\ }\textbf {\bibinfo {volume} {126}},\ \bibinfo {pages}
  {032502} (\bibinfo {year} {2021})},\ \Eprint
  {http://arxiv.org/abs/2012.13802} {arXiv:2012.13802 [nucl-ex]} \BibitemShut
  {NoStop}%
\bibitem [{\citenamefont {Koszor\'us}\ \emph {et~al.}(2021)\citenamefont
  {Koszor\'us} \emph {et~al.}}]{Koszorus:2020mgn}%
  \BibitemOpen
  \bibfield  {author} {\bibinfo {author} {\bibfnamefont {A.}~\bibnamefont
  {Koszor\'us}} \emph {et~al.},\ }\href {\doibase 10.1038/s41567-020-01136-5}
  {\bibfield  {journal} {\bibinfo  {journal} {Nature Phys.}\ }\textbf {\bibinfo
  {volume} {17}},\ \bibinfo {pages} {439} (\bibinfo {year} {2021})},\ \bibinfo
  {note} {[Erratum: Nature Phys. 17, 539 (2021)]},\ \Eprint
  {http://arxiv.org/abs/2012.01864} {arXiv:2012.01864 [nucl-ex]} \BibitemShut
  {NoStop}%
\bibitem [{\citenamefont {Neal}(1996)}]{Neal}%
  \BibitemOpen
  \bibfield  {author} {\bibinfo {author} {\bibfnamefont {R.~M.}\ \bibnamefont
  {Neal}},\ }\href@noop {} {\emph {\bibinfo {title} {{Bayesian Learning of
  Neural Networks}}}}\ (\bibinfo  {publisher} {Springer, New York},\ \bibinfo
  {year} {1996})\BibitemShut {NoStop}%
\bibitem [{\citenamefont {Hornik}\ \emph {et~al.}(1989)\citenamefont {Hornik},
  \citenamefont {Stinchcombe},\ and\ \citenamefont {White}}]{HORNIK1989359}%
  \BibitemOpen
  \bibfield  {author} {\bibinfo {author} {\bibfnamefont {K.}~\bibnamefont
  {Hornik}}, \bibinfo {author} {\bibfnamefont {M.}~\bibnamefont {Stinchcombe}},
  \ and\ \bibinfo {author} {\bibfnamefont {H.}~\bibnamefont {White}},\ }\href
  {\doibase https://doi.org/10.1016/0893-6080(89)90020-8} {\bibfield  {journal}
  {\bibinfo  {journal} {Neural Networks}\ }\textbf {\bibinfo {volume} {2}},\
  \bibinfo {pages} {359} (\bibinfo {year} {1989})}\BibitemShut {NoStop}%
\bibitem [{\citenamefont {Stone}(2013)}]{James}%
  \BibitemOpen
  \bibfield  {author} {\bibinfo {author} {\bibfnamefont {J.~V.}\ \bibnamefont
  {Stone}},\ }\href@noop {} {\emph {\bibinfo {title} {{Bayes' Rule: A Tutorial
  Introduction to Bayesian Analysis}}}}\ (\bibinfo  {publisher} {Sebtel
  Press},\ \bibinfo {year} {2013})\BibitemShut {NoStop}%
\bibitem [{\citenamefont {Casten}\ \emph {et~al.}(1987)\citenamefont {Casten},
  \citenamefont {Brenner},\ and\ \citenamefont {Haustein}}]{Casten:1987zz}%
  \BibitemOpen
  \bibfield  {author} {\bibinfo {author} {\bibfnamefont {R.~F.}\ \bibnamefont
  {Casten}}, \bibinfo {author} {\bibfnamefont {D.~S.}\ \bibnamefont {Brenner}},
  \ and\ \bibinfo {author} {\bibfnamefont {P.~E.}\ \bibnamefont {Haustein}},\
  }\href {\doibase 10.1103/PhysRevLett.58.658} {\bibfield  {journal} {\bibinfo
  {journal} {Phys. Rev. Lett.}\ }\textbf {\bibinfo {volume} {58}},\ \bibinfo
  {pages} {658} (\bibinfo {year} {1987})}\BibitemShut {NoStop}%
\bibitem [{\citenamefont {Casten}\ and\ \citenamefont
  {Zamfir}(1996)}]{Casten_1996}%
  \BibitemOpen
  \bibfield  {author} {\bibinfo {author} {\bibfnamefont {R.~F.}\ \bibnamefont
  {Casten}}\ and\ \bibinfo {author} {\bibfnamefont {N.~V.}\ \bibnamefont
  {Zamfir}},\ }\href {\doibase 10.1088/0954-3899/22/11/002} {\bibfield
  {journal} {\bibinfo  {journal} {J. Phys. G}\ }\textbf {\bibinfo {volume}
  {22}},\ \bibinfo {pages} {1521} (\bibinfo {year} {1996})}\BibitemShut
  {NoStop}%
\bibitem [{\citenamefont {Bayram}\ \emph {et~al.}(2013)\citenamefont {Bayram},
  \citenamefont {Akkoyun}, \citenamefont {Kara},\ and\ \citenamefont
  {Sinan}}]{Bayram:2013jua}%
  \BibitemOpen
  \bibfield  {author} {\bibinfo {author} {\bibfnamefont {T.}~\bibnamefont
  {Bayram}}, \bibinfo {author} {\bibfnamefont {S.}~\bibnamefont {Akkoyun}},
  \bibinfo {author} {\bibfnamefont {S.~O.}\ \bibnamefont {Kara}}, \ and\
  \bibinfo {author} {\bibfnamefont {A.}~\bibnamefont {Sinan}},\ }\href
  {\doibase 10.5506/APhysPolB.44.1791} {\bibfield  {journal} {\bibinfo
  {journal} {Acta Phys. Polon. B}\ }\textbf {\bibinfo {volume} {44}},\ \bibinfo
  {pages} {1791} (\bibinfo {year} {2013})}\BibitemShut {NoStop}%
\end{thebibliography}%

\end{document}